\documentclass[aps,preprint,prd,showpacs,nofootinbib]{revtex4}

\usepackage{amsmath,amssymb,tabularx}
\usepackage{graphicx,subfigure}
\usepackage{color}
\usepackage{booktabs}

\usepackage[colorlinks,linkcolor=magenta,anchorcolor=cyan,citecolor=blue]{hyperref}

\def\lf{\left}
\def\rt{\right}

\def\be{\begin{equation}}
\def\ee{\end{equation}}
\def\ba{\begin{eqnarray}}
\def\ea{\end{eqnarray}}

\begin{document}

\title{Towards supermassive primordial black holes from inflationary bubbles}

    \author{Hai-Long Huang$^1 $\footnote{\href{huanghailong18@mails.ucas.ac.cn}{huanghailong18@mails.ucas.ac.cn}}}


    \author{Yun-Song Piao$ ^{1,2,3,4} $\footnote{\href{yspiao@ucas.ac.cn}{yspiao@ucas.ac.cn}}}

\affiliation{$^1$ School of Physical Sciences, University of
Chinese Academy of Sciences, Beijing 100049, China}

\affiliation{$^2$ School of Fundamental Physics and Mathematical
    Sciences, Hangzhou Institute for Advanced Study, UCAS, Hangzhou
    310024, China}

\affiliation{$^3$ International Center for Theoretical Physics
    Asia-Pacific, Beijing/Hangzhou, China}

\affiliation{$^4$ Institute of Theoretical Physics, Chinese
    Academy of Sciences, P.O. Box 2735, Beijing 100190, China}

\begin{abstract}

The bubbles that nucleated during slow-roll inflation can be
supercritical, i.e. their radii are larger than the Hubble horizon
of de Sitter spacetime inside the bubble (an inflating baby
universe inside it), and thus naturally develop to the
supermassive primordial black holes (SMPBHs) with a multi-peaks
mass function. In this paper, we further investigate relevant
phenomenology. After slow-roll inflation ended, the bubbles may be
not only supercritical, but also subcritical. It is showed that it
seems unlikely for the subcritical bubbles to collapse to SMPBHs.
Theoretically, however, before they collapsed such bubbles might
have a probability of up-tunnelling to the supercritical ones and
thus contribute to SMPBHs. We present a mechanism for the origin
of initial clustering of SMPBHs, which can significantly magnify
the merger rate of SMPBH binaries, and show the possibility that
the merging of such SMPBH binaries explains recent NANOGrav
signal.

\end{abstract}

    \maketitle
    \tableofcontents


\section{Introduction}

Recently, lots of observational hints for supermassive black holes
(SMBHs) have been gathered. It has been believed that $M\sim
10^6-10^{9}M_\odot$ SMBHs occupied at the center of galaxies
observed at redshifts $z\gtrsim 6$. And at lower redshift, the
merger of SMBHs with $M\gtrsim 10^9M_\odot$ might be the source of
nano-Hertz gravitational wave (GW) background recently detected by
PTA experiments
\cite{NANOGrav:2023gor,Xu:2023wog,Reardon:2023gzh,EPTA:2023fyk},
while at higher redshift $z\sim 10$, the observations with JWST
have discovered lots of early supermassive galaxies ($M\gtrsim
10^{10}M_\odot$), which seems to be incompatible with the
well-known $\Lambda$CDM model, but is likely to be solved with
$\Lambda$CDM+SMBHs($M\gtrsim 10^9M_\odot$)
\cite{Liu:2022bvr,Hutsi:2022fzw,Huang:2023chx}.

However, \textsf{how such SMBHs can form} has still been a
challenge to the standard astrophysical accretion
models~\cite{Volonteri:2010wz,Volonteri:2021sfo}. Though SMBHs can
come into being by the accretion of seed BHs with $\sim
10^3M_\odot$ ~\cite{Duechting:2004dk,Serpico:2020ehh}, but the
accretion might be not sufficiently efficient, especially for
SMBHs ($\gtrsim 10^9M_\odot$) observed at redshift $z>10$. Thus it
might be the most natural that such SMBHs are primordial black
holes (PBHs), actually current observations did not rule out the
possibility that SMPBHs has a density fraction, $f_{\text{PBH}}\lesssim
10^{-3}$, see e.g.\cite{Carr:2020erq}.

In string landscape with metastable vacua
\cite{Bousso:2000xa,Susskind:2003kw,Kachru:2003aw} (usually
modelled as a multi-dimensional potential), the slow-roll
inflation might happen at certain metastable state, and the
bubbles with different vacua will inevitably nucleate in
corresponding inflating region \footnote{It is possible that the
bubbles might also arise naturally in models with large adiabatic
fluctuations, e.g.\cite{Escriva:2023uko,Atal:2019erb}, which can
happen in the case of single field slow-roll potential with a
small barrier.}. It has been found in Ref.\cite{Garriga:2015fdk}
that in such a scenario \footnote{This is a reminiscent of
well-known eternally inflating multiverse
\cite{Vilenkin:1983xq,Linde:1986fd}, see also
Refs.\cite{Guth:2007ng,Linde:2015edk}.}, after slow-roll inflation
ended, both the subcritical and supercritical bubbles
\footnote{Here, the critical bubble refers that its radius equals
to the Hubble horizon of de Sitter spacetime inside bubble, in
light of Ref.\cite{Garriga:2015fdk}.} might evolve into PBHs,
while the supercritical bubbles (with a baby inflating universe
inside it) will possibly contribute to SMPBHs, see also
\cite{Deng:2016vzb,Deng:2017uwc,Wang:2018cum,Liu:2019lul,Deng:2020mds,
Ashoorioon:2020hln,Ashoorioon:2022raz} for subsequent
investigations. However, it has been showed in
Refs.\cite{Garriga:2015fdk,Deng:2017uwc} that the mass
distribution of such multiverse PBHs is $\propto {1\over
M^{1/2}}$, see also \cite{Kusenko:2020pcg}, which thus is
negligible at supermassive band $M\gtrsim 10^{9}M_\odot$.

In past years, the cosmological implications of PBHs
\cite{Zeldovich,Hawking:1971ei,Carr:1974nx} have been intensively
studied, see relevant reviews
\cite{Sasaki:2018dmp,Carr:2020gox,Carr:2023tpt,Domenech:2023jve}.
It has been observed earlier in
Refs.\cite{Carr:1993aq,Ivanov:1994pa,Garcia-Bellido:1996mdl,Kawasaki:1997ju,Yokoyama:1998pt}
that
large inflationary perturbations ($\delta\rho/\rho\gtrsim 0.1$)
can source PBHs \footnote{Recently, lots of the mechanisms of
making PBHs have been presented, related to inflation
e.g.\cite{Inomata:2017okj,Kannike:2017bxn,Cheng:2018yyr,Lin:2020goi,
Ashoorioon:2020hln,Cai:2021wzd,Kawai:2021edk,Karam:2022nym,Papanikolaou:2022did,
Fu:2022ssq,Garcia-Bellido:2017mdw,Germani:2017bcs,Byrnes:2018txb,Fu:2019ttf,
Fu:2019vqc,Fu:2020lob,Zhai:2023azx,Ragavendra:2020sop,Fu:2022ypp,Di:2017ndc,
Motohashi:2017kbs,Yi:2020cut,Ballesteros:2018wlw,Kamenshchik:2018sig,Qiu:2022klm,
Cai:2019bmk,Nakama:2018utx,
ZhengRuiFeng:2021zoz,Gao:2020tsa,Lin:2021vwc,Pi:2021dft,Domenech:2021wkk,DeLuca:2022bjs,
Kawaguchi:2023mgk,Choudhury:2023hfm,Choudhury:2023fwk,Domenech:2023dxx,Pi:2022zxs,
Meng:2022ixx,Cai:2023uhc}, or
\cite{Liu:2021svg,He:2022amv,Lewicki:2023ioy}.}. However, current
CMB spectral distortion observations have ruled out such a
significant enhancement of the perturbation amplitudes on
$k\lesssim 10^4$Mpc$^{-1}$ scale, so $M
> 10^4M_\odot$ PBHs
\cite{Nakama:2017xvq}, see also
Refs.~\cite{DeLuca:2021hcf,DeLuca:2022bjs}. It seems that one
needs a scenario with highly non-Gaussian primordial perturbations
to create SMPBHs, e.g.\cite{Nakama:2016kfq,Atal:2020yic} and
recent \cite{Hooper:2023nnl}.


However, the scenario of PBHs from inflationary bubbles is
not constrained by such CMB observations. It has been showed in
Ref.\cite{Huang:2023chx} that when the inflaton slowly passes by a
neighboring vacuum, the nucleating rate of supercritical bubbles
would inevitably present a peak, thus the resulting PBHs not only
can be naturally supermassive, $M\gtrsim 10^5M_\odot$, or even
$10^{9}M_\odot$, but also have a non-negligible peak-like
mass distribution. However, the phenomenology of such bubbles to
SMPBHs might be more colorful than expected. The initial
clustering of PBHs can significantly magnify the merger rate of
PBH binaries \footnote{The initial clustering of PBHs and its
implication has also been studied in
Ref.\cite{Chisholm:2005vm,DeLuca:2022bjs,Ali-Haimoud:2018dau,
Desjacques:2018wuu,Ballesteros:2018swv,Belotsky:2018wph,
Bringmann:2018mxj,Suyama:2019cst,Young:2019gfc,Shinohara:2021psq,
DeLuca:2021hcf,DeLuca:2021hde,Kawasaki:2021zir,Kasai:2023ofh} .},
which might be necessary for the detectability of SMPBHs merger,
e.g.\cite{Ding:2019tjk,Depta:2023qst,Gouttenoire:2023nzr}. Is
initial SMPBHs clustering possible? After slow-roll
inflation ended, the bubbles may be not only supercritical, but
also subcritical. It is also natural and interesting to ask
whether the subcritical bubbles can collapse into SMPBHs or not.

In section-\ref{section_review}, we briefly review how the
supercritical bubbles evolve to SMPBHs. In
section-\ref{section_subcritical}, we focus on the subcritical
bubbles, and show that it seems unlikely for such bubbles to
collapse into SMPBHs. However, we find that the subcritical bubble
might have a probability of up-tunnelling to the supercritical one
and thus contribute to SMPBH. In section-\ref{section_cluster}, we
present a mechanism for the origin of initial clustering of
SMPBHs, and show its implications for the merger rate of SMPBHs
and recent NANOGrav signals.



\section{Review on SMPBHs sourced by supercritical bubbles}
\label{section_review}

\begin{figure}[htbp]
    \centering
    \includegraphics[width=3.8in]{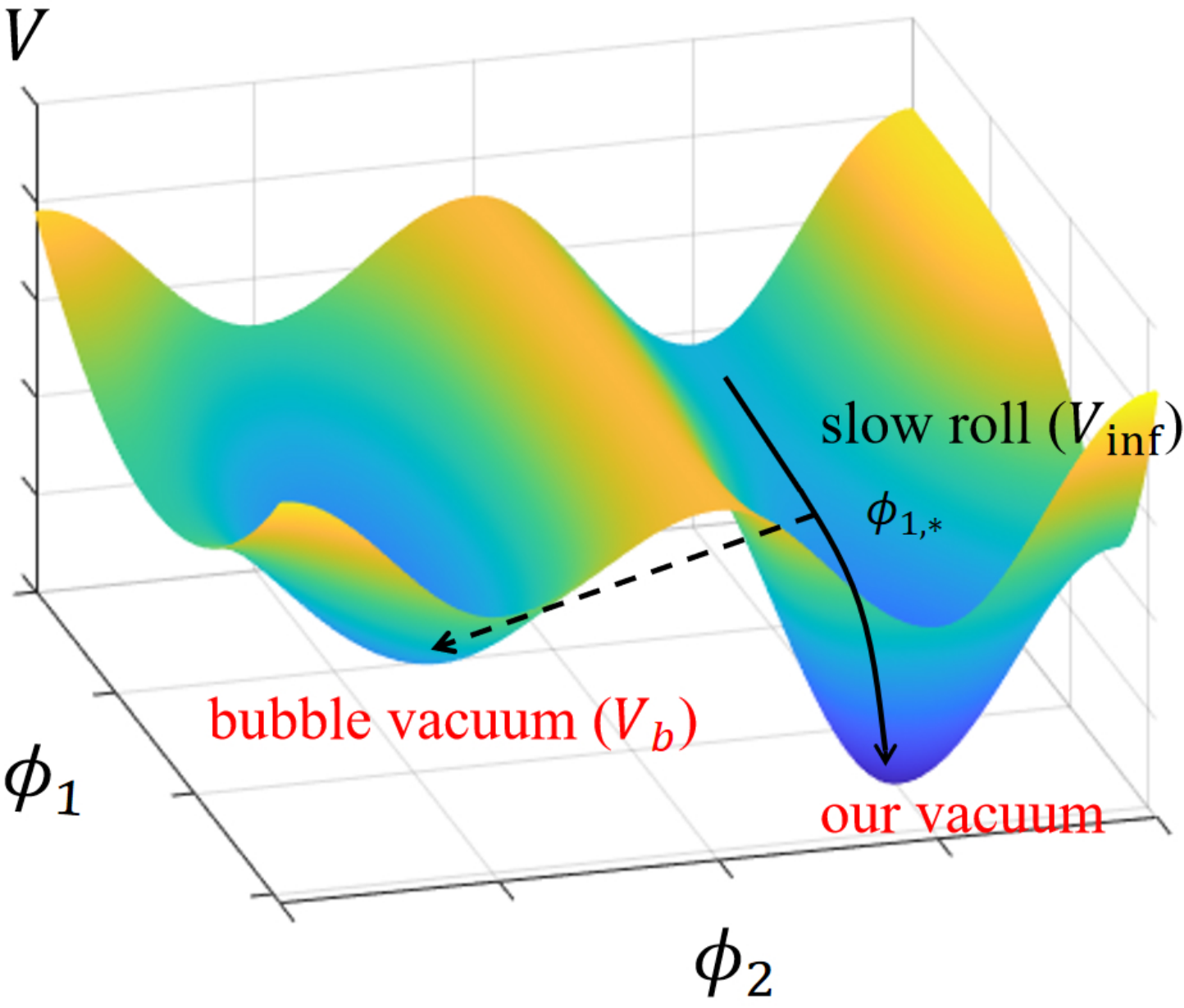}
\caption{\textbf{A 2D potential for our model.} Initially, the
slow-roll inflaton $\phi_1$ rolls along its potential $V_{\text{inf}}$ at
$\phi_2=0$, with a neighboring vacuum $V_b<V_{\text{inf}}$. In our setup,
only when inflaton rolls to $\phi_1=\phi_{1,*}$, the corresponding
vacuum bubbles will just nucleate with the largest rate. }
    \label{Two-FieldPotential}
\end{figure}

In our phenomenological model \cite{Huang:2023chx}, see
\autoref{Two-FieldPotential}. Initially, the inflaton $\phi_1$
slowly rolls along its potential $V_{\text{inf}}(\phi_1)$. The nucleating
rate of bubble with the vacuum $V_b<V_{\text{inf}}$ has a peak at
$\phi_1=\phi_{1,*}$ \footnote{The effect of the rolling velocity of
inflaton on the nucleating rate, so the mass distribution of PBHs,
has been calculated in Ref.\cite{Huang:2023chx,Huang:2023klk}, see
also \cite{Kleban:2023ugf} for a different perspective. Here we
will not involve it.}, i.e. the efolds number of inflation ${\cal
N}={\cal N}_*$.

The bubble after its nucleating rapidly expands, and after its
radius exceeds the Hubble horizon of slow-roll inflating
background, i.e. $r>1/H_i$, where $H^2_i={8\pi\over
3M_P^2}V_{\text{inf}}$, it will expand comovingly with cosmological
background. The inflation ended at $t_i$, at which
\be 
r_i={e^{{\cal N}_*}\over H_i}\gg 1/H_i,\label{Nstar}\ee and the
energy of inflaton is rapidly converted to that of radiation, so
though $V=V_b$ inside the bubble, $V\sim 0$ outside the bubble is
completely negligible.

The supercritical bubble refers that when inflation ended its
radius exceeds the Hubble horizon of de Sitter spacetime inside
bubble, $r_i>1/H_b$, thus we have \be {\cal N}_*>\ln{H_i\over
H_b}.\ee where $H^2_b={8\pi\over 3M_P^2}V_b$. This suggests that
its interior contains a baby inflating universe, which is
connected to the exterior (our observable Universe) through a
wormhole, see Appendix-\ref{section_critical}, see also
Refs.~\cite{Maeda:1981gw,Kodama:1981gu} \footnote{The possibility
of creating a baby universe in the laboratory (asymptotically flat
spacetime) has been explored in seminal
Refs.\cite{Farhi:1986ty,Farhi:1989yr}.}.

\begin{figure}[t]
   \includegraphics[width=10cm]{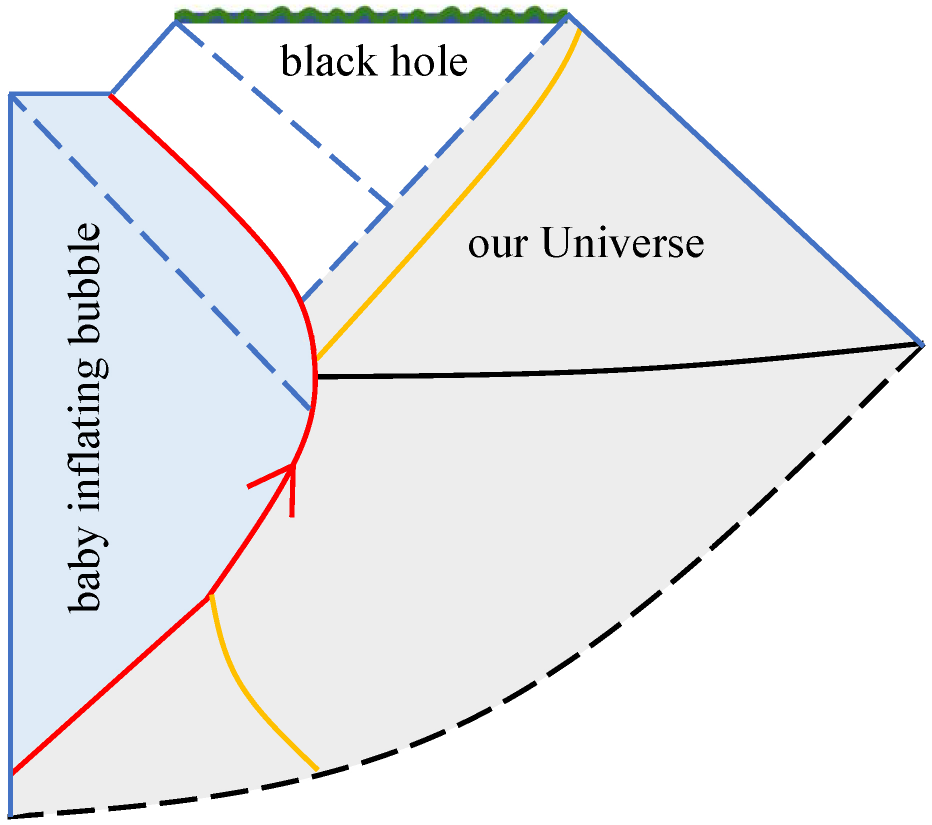}
\caption{\textbf{Causal diagram of supercritical bubble to PBH
\cite{Huang:2023chx}.} Inflation ended at $t=t_i$ (black solid
curve). The red and yellow curves correspond to the comoving
bubble wall and Hubble horizon, respectively. The interior of
supercritical bubble contains a baby inflating universe, since
$r_i\gtrsim 1/H_b$. The bubble wall expanded comovingly with its
exterior spacetime, and the bubble will be hidden behind the
horizon of a BH shortly after it entered into the Hubble horizon,
$r\sim {1/ H}$. }
    \label{fig:Penrose1}
\end{figure}

In radiation era the bubble expanded comovingly with the scale factor
$a(t)=({t\over t_i})^{1/2}$, and its radius will arrive at the
cosmological horizon at $t_H$, after which it will be hidden
behind the horizon of a PBH, see \autoref{fig:Penrose1}, with mass
$M_{\text{PBH}}\sim M_P^2t_{H}$, thus we have \ba M_{\text{PBH}}\simeq
{M_P^2\over H_i}e^{2{\cal N}_*}.\label{Msuper}\ea
Provided that the supercritical bubbles nucleated at ${\cal
N}_*\gtrsim {1\over 2}\ln{10^9M_\odot\over M_P^2/H_i}\simeq 48$
(the corresponding wavenumber is $k_*\lesssim 100\text{Mpc}^{-1}$),
we will naturally have $M_{\text{PBH}}\gtrsim 10^9M_\odot$ in our Universe. In
models that large inflationary perturbations source PBHs, the
amplitude of scalar perturbation must satisfy $P_s\sim {H^2\over
\epsilon M_P^2}\gtrsim 0.01$ \footnote{In single field
ultra-slow-roll inflation, the large perturbations at PBH scales
might bring unacceptable loop corrections to the perturbations at
CMB scales
\cite{Kristiano:2022maq,Riotto:2023hoz,Choudhury:2023vuj,Kristiano:2023scm,Riotto:2023gpm,Choudhury:2023rks,Choudhury:2023kdb}.}
at PBH scales, which is conflicted with the CMB spectral
distortion observation. How to avoid this confliction has been
still a significant challenge. Here, the primordial perturbations
on all scales outside the bubbles are set by slow-roll inflation,
see \autoref{Two-FieldPotential}, thus we always have $P_s\ll
0.01$.

In the context of $\Lambda$CDM, we have \be {\cal
N}_{\text{eq}}=\ln{a(t_{i})H_{i}\over
a_{\text{eq}}H_{\text{eq}}}\thickapprox \ln\sqrt{H_i\over
H_{\text{eq}}},\label{Nstar}\ee where ${\cal N}_{\text{eq}}$
corresponds to the efold number at matter-radiation equality,
noting $a(t)=({t\over t_i})^{1/2}$ for the radiation era. It is
usually required that ${\cal N}_*\lesssim {\cal N}_{\text{eq}}$.
According to \autoref{Msuper}, we have the upper bound for SMPBHs
\be M_{\text{PBH}} < {M_P^2\over H_i}e^{2{\cal
N}_{\text{eq}}}={M_P^2\over H_{\text{eq}}}\sim 10^{18}M_\odot. \ee
It has been also pointed out in Ref.\cite{Huang:2023chx} that the
slow-roll path of inflaton might be accompanied with more than one
neighboring vacua, thus the corresponding SMPBHs might present a
multi-peaks mass distribution at different mass bands below
$10^{18}M_\odot$, see also section-\ref{section_cluster}.

\section{On subcritical bubbles and PBHs}
\label{section_subcritical}

\subsection{Can subcritical bubbles collapse to SMPBHs ?}\label{sec:three_A}

\begin{figure}[t]
   \includegraphics[width=10cm]{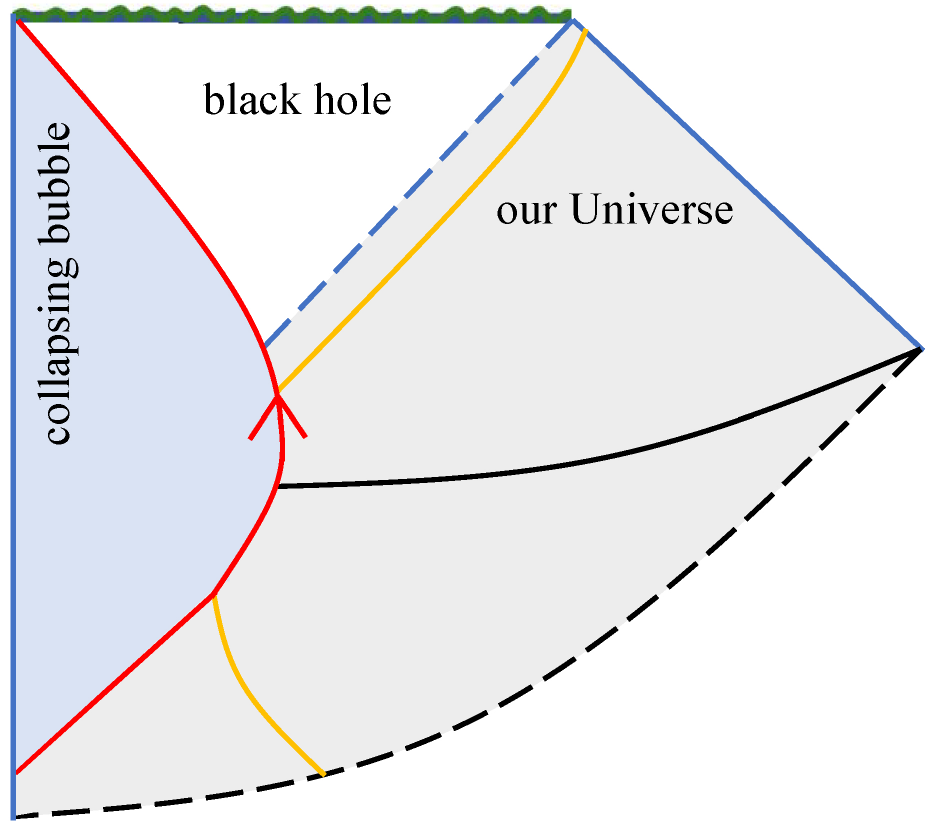}
\caption{\textbf{Causal diagram of subcritical bubble to PBH.} The
red and yellow curves correspond to the comoving bubble wall and
Hubble horizon, respectively. The subcritical bubble ($r_i<
1/H_b$), shortly after inflation ended (black solid curve), will
stop expanding, since the motion of bubble is bounded by a
``barrier", see Appendix-\ref{section_critical}, and inevitably
collapse.}
    \label{fig:Penrose2}
\end{figure}

The bubble might be also subcritical (its radius is smaller than
the Hubble horizon of spacetime inside bubble when inflation
ended) \footnote{The corresponding bubble generally
nucleated at later period (smaller ${\cal N}_*$) of slow-roll
inflation. In this case, it is necessary that $H_b<H_i$. },
$r_i<1/H_b$. In this case, we have \be {\cal N}_*<\ln{H_i\over
H_b}.\ee However, unlike the supercritical bubble, since the
velocity of bubble wall is slowed down by the negative pressure
pulled inwards, the subcritical bubble will stop expanding shortly
after inflation ended, and inevitably collapse towards the
Schwarzschild singularity \cite{Garriga:2015fdk}, see
\autoref{fig:Penrose2} and also Appendix-\ref{section_critical}.

The metric inside the interior of bubble is the de Sitter metric,
while that outside the bubble wall is the Schwarzschild-like
metric. In thin-wall approximation the mass of subcritical bubble
is \cite{Berezin:1982ur,Blau:1986cw} \footnote{Here,
$\sigma=\int_{path}
\sqrt{2V_b}\lf(\sum_{i=1,2}\text{d}\phi_i^2\rt)^{1/2}$ is the wall
tension of bubble, with the ``path" representing the
``least-$\sigma$" path in multiple-field potential in
\autoref{Two-FieldPotential}, see Ref.\cite{Huang:2023chx}.} \ba
\label{massparameter} M_{\text{bubble}} = {4\pi\over 3}r^3V_b+{4\pi r^2\sigma}
\sqrt{1+\lf({\text{d}r\over \text{d}\tau}\rt)^2-H_b^2r^2}-8\pi^2{\sigma^2
r^3\over M_P^2}.  \ea The radius of subcritical bubble is bounded
by a barrier $V(r)$, see Appendix-\ref{section_critical}. Thus it
is expected that shortly after inflation ended the bubble wall
will meet the barrier (at $r_{\text{cls}}$), the expansion of bubble will
end and its collapse starts. Thus when the bubble started to
collapse, ${\text{d}r_{\text{cls}}\over \text{d}\tau}=0$, with negligible
$\sigma$ \footnote{This corresponds to $\sigma\ll V_b r_{\text{cls}}$
and $\sigma\ll V_b^{1/2}M_P$.}, we have \ba M_{\text{bubble}}\sim {4\pi\over
3}r_{\text{cls}}^3V_b. \label{subPBH}\ea

It is thought in Ref.\cite{Garriga:2015fdk} that the bubble mass
is constant in collapsing era of bubble. In this case, we have \be
M_{\text{PBH}}=M_{\text{bubble}}\sim {4\pi\over 3}r_{\text{cls}}^3V_b <{M_P^2\over
2H_i}e^{{\cal N}_*}\lf({r_{\text{cls}}\over
r_i}\rt)^3.\label{eq:Msubcritical}\ee Actually, when inflation
ended, $M_{\text{bubble}}\thickapprox {1\over 2}H_b^2M_P^2r_i^3\sim
{1\over 2}H_b^2M_P^2e^{3{\cal N}_*}H_i^{-3}$. However, since
$H_br_i<1$ for subcritical bubble, we have
\autoref{eq:Msubcritical}. Thus the mass of PBHs that such
subcritical bubbles collapse to is \be M_{\text{PBH}}< {M_P^2\over
2H_i^{1/2}H_{\text{eq}}^{1/2}}\lf({r_{\text{cls}}\over r_i}\rt)^3\sim
10^{-11}\lf({M_P\over H_i}\rt)^{1/2}\lf({r_{\text{cls}}\over
r_i}\rt)^3M_\odot. \label{subM1}\ee The scale that slow-roll
inflation happened might be $M_{\text{inf}}\sim 10^{15}$Gev, thus we have
\be M_{\text{PBH}}<10^{-11}{M_P\over M_{\text{inf}}}\lf({r_{\text{cls}}\over r_i}\rt)^3
M_\odot\sim 10^{-7}\lf({r_{\text{cls}}\over r_i}\rt)^3 M_\odot.\ee In this
sense, it seems that the subcritical bubbles can hardly collapse
to SMPBHs, unless ${r_{\mathrm{cls}}\over r_i}>10^{4}$. However, slow-roll
inflation might happen at a lower scale than expected. According
to \autoref{subM1}, such PBHs can be supermassive, $M_{\text{PBH}}\gtrsim
10^5M_\odot$, only when inflation happened at $M_{\text{inf}}\lesssim
10^3$Gev ($H_i^{1/2}M_P^{1/2}\lesssim 10^3$Gev).




\subsection{General-relativistic collapse of subcritical
bubble}\label{section_theory}

In subsection-\ref{sec:three_A}, it is supposed that
$M_{\text{PBH}}=M_{\text{bubble}}$ (the mass of subcritical bubble
when the bubble started to collapse). However, unlike the
supercritical bubble that always expands, the collapse of
subcritical bubble will excite the large inhomogeneity. Thus it is
significant to inspect the effects of such non-perturbative
collapse on $M_{\text{PBH}}=M_{\text{bubble}}$. In this
subsection, we will follow Refs.\cite{Lin:2021ubu,Lin:2022ygd},
and perform the 3+1D numerical relativity (NR) \footnote{In the
past decades, NR has been developed significantly
\cite{Pretorius:2005gq,Campanelli:2005dd,Baker:2005vv}, which has
been applied to the studies on non-perturbative cosmologies,
e.g.\cite{Giblin:2015vwq,Bentivegna:2015flc,East:2015ggf,Clough:2016ymm,deJong:2021bbo}.}
simulation for the evolution of subcritical bubbles.

\begin{figure}[htbp]
    \centering
    \includegraphics[width=3.1in]{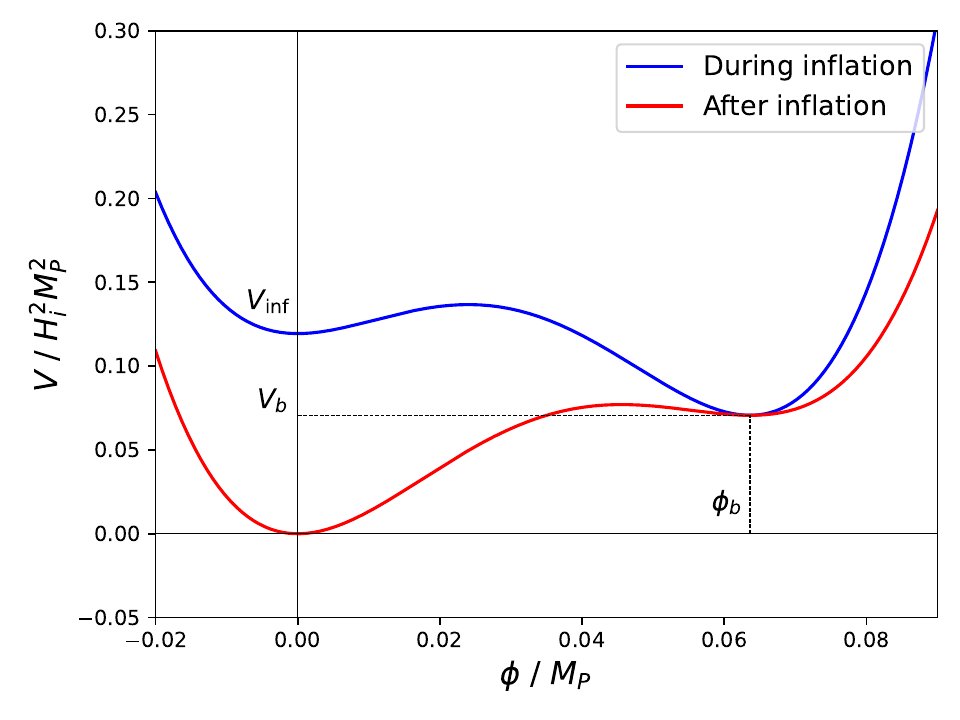}
\caption{\textbf{The potentials of $\phi$ before and after
inflation ended (the blue and red curves), respectively.} In our
simulation, it is assumed that after inflation ended, inside the
bubble $V_b=const.$ is unchanged, while the energy
of inflaton outside the bubble is convert rapidly into dust or
radiation, so $V_{\text{inf}}\sim 0$.}
    \label{V2}
\end{figure}

It is convenient to consider a 1D effective potential in
\autoref{V2},
\begin{gather}
V={1\over 2} \lf(m^2-\frac{\gamma^2}{2\lambda}\rt)
\phi^2+ \lambda\phi^2\left(\phi-{\gamma\over
2\lambda}\right)^2+V_{\text{inf}},  \label{V1}
    \end{gather}
instead of (as a simplification of) \autoref{Two-FieldPotential},
where the inflation happened at $\phi=0$ ($V=V_{\text{inf}}\simeq
const.$). In our setup, when inflation ended, inflaton is rapidly
converted to matter (dust or radiation), thus outside the bubble
$V=V_{\text{inf}}\sim 0$ is completely negligible and the initial density
of matter is $\rho\sim V_{\text{inf}}$.
The post-inflation potential barrier can be significantly
different, see \autoref{Two-FieldPotential}. Here, for simplicity,
we set it still described by \autoref{V1}, but with different
values of the relevant parameters $m,\lambda,\gamma$, see in
\autoref{V2}.

The dust or radiation is modelled with an oscillating scalar field
$\psi$ with $V\sim \psi^2$ or $ \psi^4$, as in
e.g.\cite{deJong:2021bbo}. And except the gravity, the
interactions of the field $\phi$ with $\psi$ (the surrounding
matter) is negligible.

To perform our simulation, we modify the NR package
GRChombo \footnote{http://www.grchombo.org\\https://github.com/GRChombo}
\cite{Clough:2015sqa,Andrade:2021rbd}. In the context of 3+1
decomposition, the metric is $g_{00} = -{\alpha}^2+{\beta}_i
{\beta}^i$, $g_{0i} = {\beta}_i$ and $g_{ij} = {\gamma}_{ij}$,
where $\alpha$ is the lapse parameter, ${\beta}^i$ the shift
vector and ${\gamma}_{ij}$ the spatial metric. The conformal
metric $\widetilde{\gamma}_{ij}=\chi \gamma_{ij}$, the connections
$\widetilde{\Gamma}^i= \widetilde{\gamma}^{jk}
\widetilde{\Gamma}^i_{jk}$ and the extrinsic curvature ${\cal
K}_{ij}=\frac{1}{3}{\cal K}\delta_{ij}+A_{ij}$ evolve according to
the BSSN equations
\cite{Baumgarte:1998te,Shibata:1995we} \footnote{Initially, we set
the BSSN variables $\widetilde{\gamma}_{ij}=\delta_{ij}$,
$\widetilde{A}_{ij}=\chi A_{ij}=0$, $\chi=1$ which is then relaxed
in the light of ${\partial}_t \chi =\mathcal{H}$ to satisfy the
momentum and Hamiltonian constraints, see also
e.g.\cite{Clough:2016ymm,Lin:2021ubu}. }, see
Appendix-\ref{section_NR}.


\begin{figure}[tb]
   \subfigure[] {\includegraphics[width=3.1in]{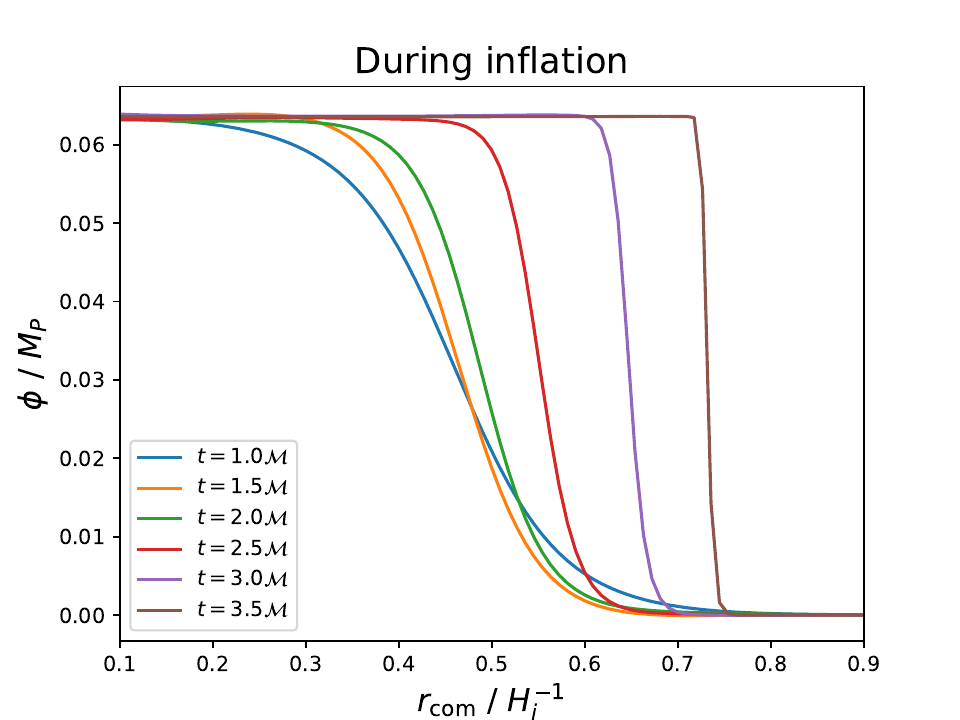}}
    \quad
   \subfigure[] {\includegraphics[width=3.1in]{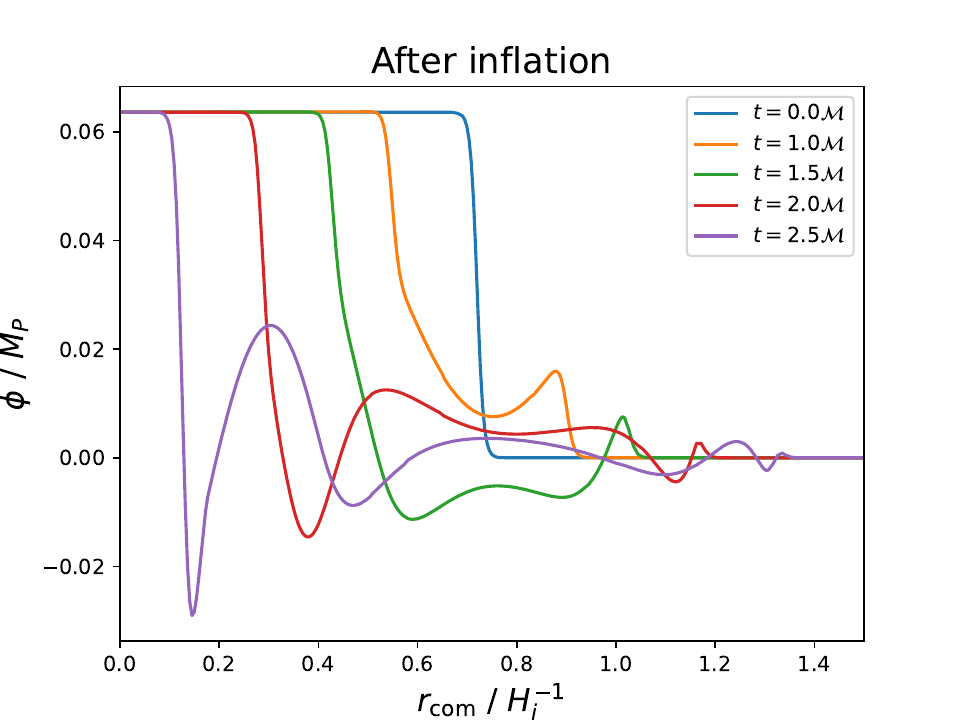}}
\caption{\textbf{Profiles of the bubbles during inflation (left
panel) and after inflation (right panel).} Here, $1{\cal
M}=0.1342/H_i$ and $r_{\text{com}}$ is the comoving radial
coordinate. In our simulation, the initial profile of bubble is
set as $\phi(t=0,x^i)=\lf[1-\text{tanh}\lf(
    \frac{r_{\text{com}}-r_{0}}{l}\rt)\rt]{\phi_b}/{2}$,
where the parameters $r_{0}$ and $l$, respectively, depict the
radius of the bubble and the width of bubble wall, then the
profile of bubble is solved with NR until inflation ended, and the
results at the time when inflation ended are taken as the initial
conditions of the post-inflation evolution of bubble.}
    \label{result}
\end{figure}

\begin{figure}[htbp]
    \centering
    \includegraphics[width=6.9in]{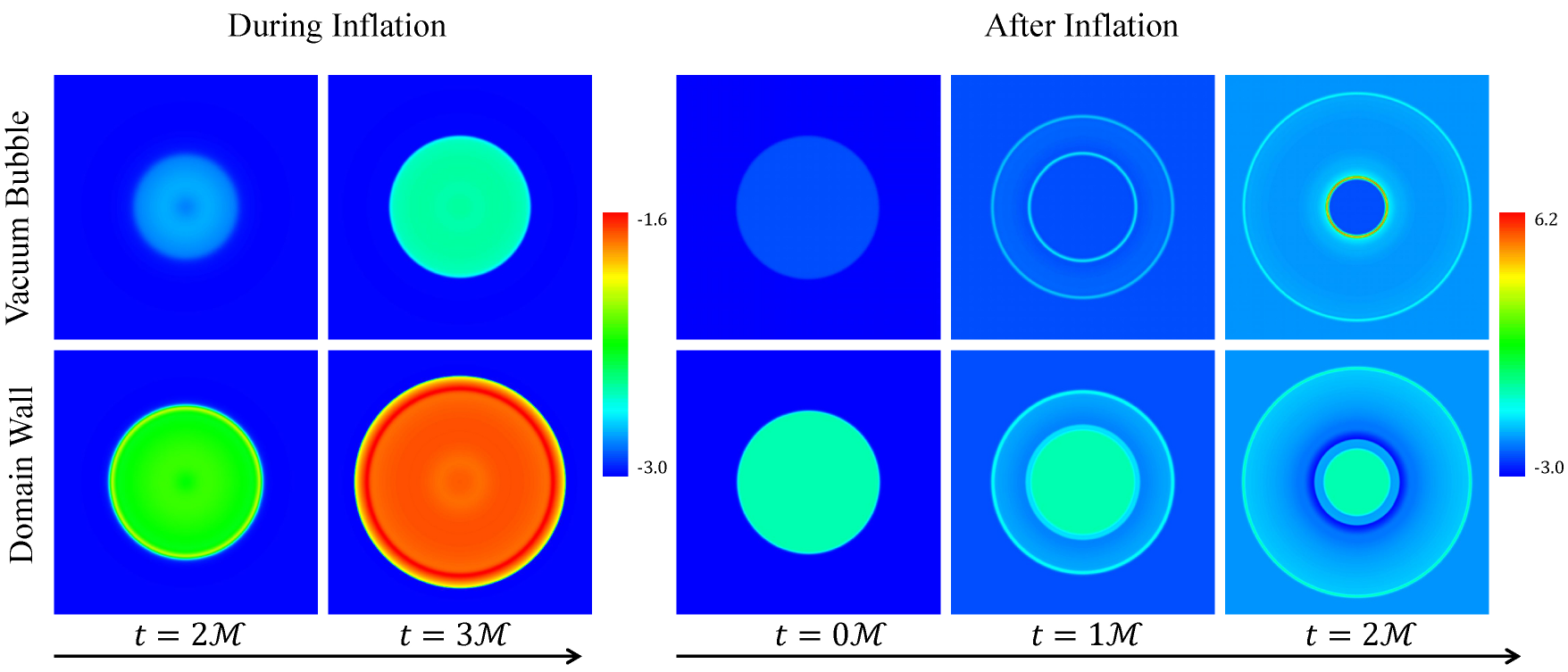}
\caption{\textbf{Snapshots of spacetime inside and outside the
bubble,} where the external curvature is in unit of $H_i$.
The upper and lower panels are for the bubbles with $V_b\neq0$ and
$V_b=0$ (in which case the bubble corresponds to a
spherical domain wall moving in a homogeneous background after
inflation). Initially, the bubble wall is expanding (the local
Hubble rate $H_{\rm local}=-{\cal K}/3>0$), however, shortly after
inflation ended, it starts to collapse ($H_{\rm local}<0$). The
collapse of bubble will excited large fluctuations of scalar field
at the regions which the bubble wall sweeps through, which makes
the field at the corresponding region oscillating drastically, see
outward wave circles, and also the right panel of
\autoref{result}. }
    \label{K}
\end{figure}

The simulation results are showed in \autoref{result} and
\autoref{K}. Initially, the bubble expands, its comoving radius
$r_{\text{com}}$ will be close to but \be
r_{\text{com}}=\int^{t\gg 1/H_ i}{\text{d}t\over e^{H_it}}<
{1/H_i},\ee and after inflation ended the bubble wall still
expanded due to inertia, but its velocity is slowed down by the
negative pressure pulled inwards, so that shortly the bubble will
inevitably collapse, as showed in
Ref.\cite{Garriga:2015fdk}.

However, unexpected, it seems that most energy of bubble will
dissipate outwards rapidly in the collapsing phase of bubble. The
collapse of bubble excited large fluctuations of scalar field on
the bubble wall, which makes the field at the corresponding region
oscillating drastically when the bubble wall sweeps through
certain region. The oscillating of field also inevitably causes
that of local spacetime, so the local expansion rates at different
regions are different, see the circles outside the bubble in
\autoref{K}. These oscillating energy will be diluted rapidly by
the expansion of spacetime outside the bubble.

\begin{figure}[tb]
    \subfigure[] {\includegraphics[width=3.1in]{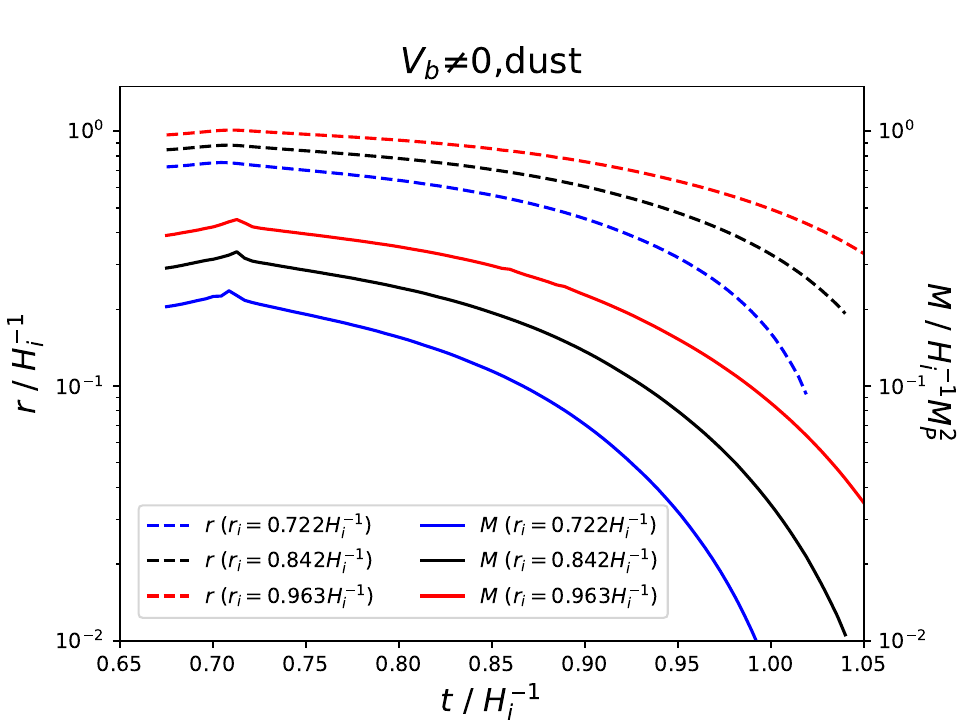}}\label{dustbubble}
    \quad
    \subfigure[] {\includegraphics[width=3.1in]{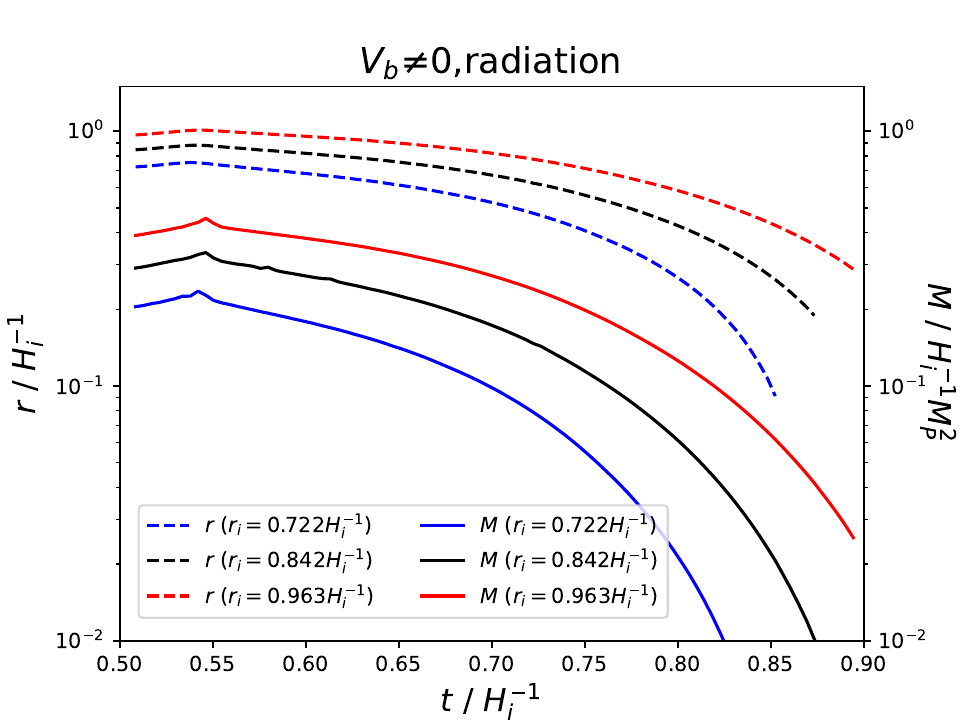}}\label{radiationbubble}
   \subfigure[] {\includegraphics[width=3.1in]{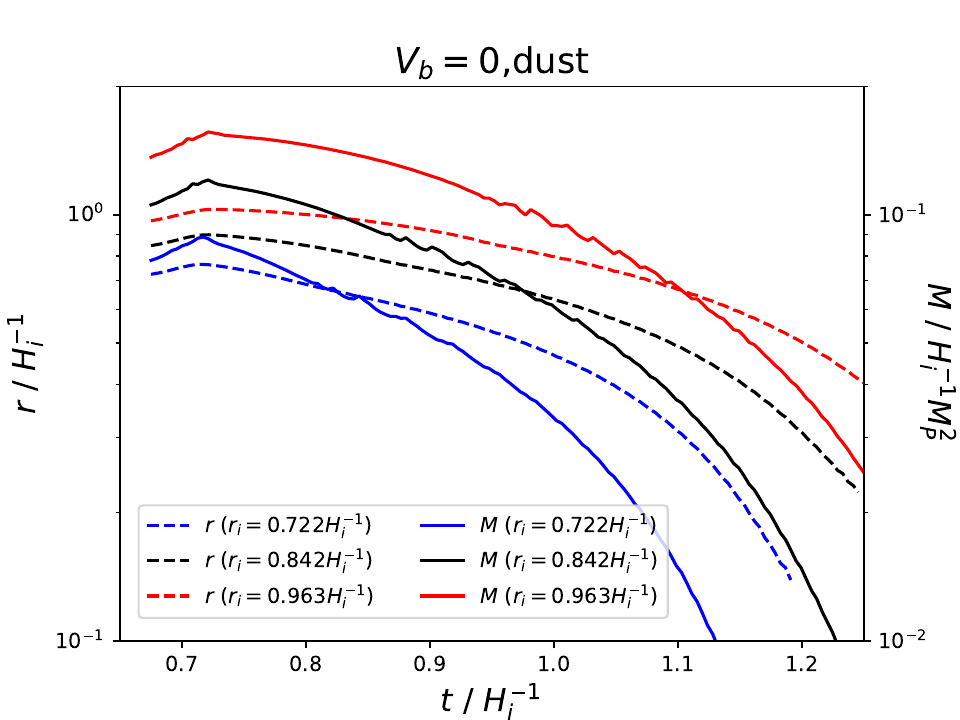}}\label{dustdomainwall}
    \quad
    \subfigure[] {\includegraphics[width=3.1in]{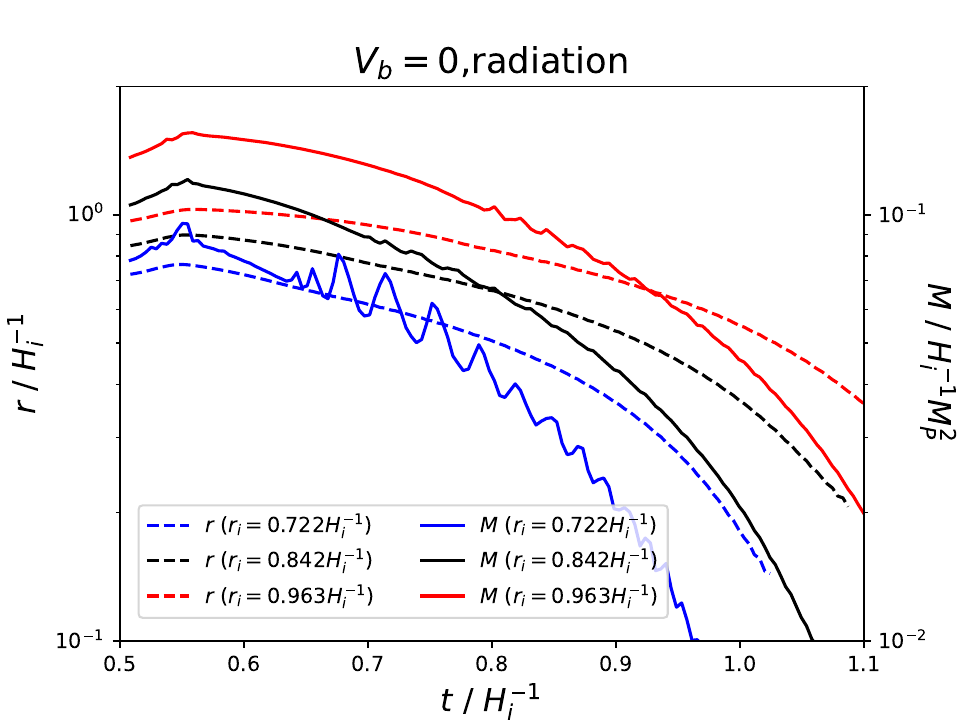}}\label{radiationdomainwall}
\caption{ \textbf{The radius and mass of the bubble with respect
to the time,} in dust and radiation-like universes, respectively,
after inflation ended. The upper and lower panels are for the
bubbles with $V_b\neq0$ and $V_b=0$. Here, we set $a_i=e^{H_i
(t_i-t_*)}=1$ at $t_i$ when inflation ended. }
    \label{massandR}
\end{figure}

According to \autoref{massparameter}, we show the mass $M$ of
bubbles after inflation ended in \autoref{massandR}
\footnote{Here, we have implemented lots of simulations with
different values of the parameters of potential barrier in
\autoref{V1}, and will send detailed results upon request.}, and
observe that it rapidly sink in collapsing phase of bubble. Though
when inflation ended the comoving radius of bubble satisfies
${r}_{\text{com},i}< {1/H_i}$, its physical radius is
$r_i={r}_{\text{com},i}e^{H_i (t_i-t_*)}\gg {1/ H_i}$. Thus our
results in \autoref{massandR} are applicable for arbitrary $r_i\gg
{1/ H_i}$. In the case that the bubble can collapse into the PBH,
we must have \footnote{This is well-known \textsf{the hoop
conjecture}. Here we ignore the effect of the presence of an
expanding background \cite{Saini:2017tsz}, see
\cite{deJong:2021bbo}. } \be r\lesssim r_{\text{Sch}}={2M\over
M_{P}^2}, \label{loop}\ee where $r_{\text{Sch}}$ is the
Schwarzschild radius of bubble.
In light of \autoref{massandR}, since $r_{\text{Sch}}$ shrinks rapidly,
it seems to be impossible for such bubbles to collapse to massive
PBHs.

Thus though further research on the NR simulation in cosmology is
required, \autoref{subM1} might overestimate the mass of PBHs.
In view of this, the mass of PBHs sourced by the subcritical
bubbles is worth surveying any further.

\subsection{Up-tunnelling of subcritical bubbles}
\label{subsection_tunnel}

\begin{figure}[t]
   \includegraphics[width=10cm]{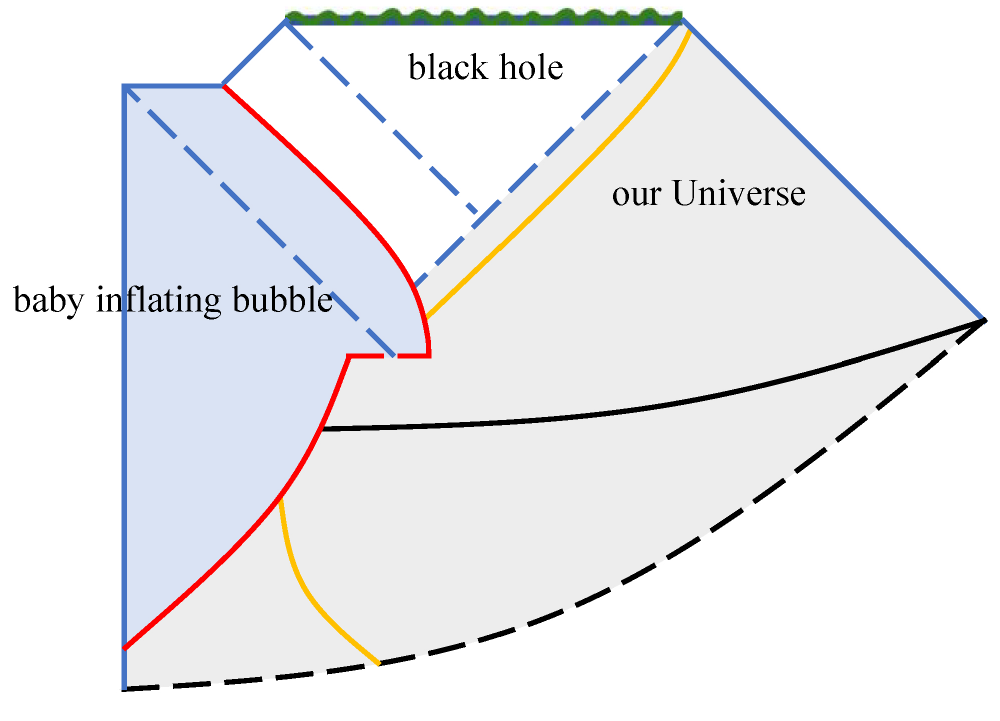}
\caption{\textbf{Causal diagram of up-tunnelling subcritical
bubble.} Inflation ended at $t=t_i$ (black solid curve). The red
and yellow curves correspond to the comoving bubble wall and
Hubble horizon, respectively. The subcritical bubble ($r_i<
1/H_b$), before its starting to collapse, has a probability of
up-tunnelling (red dashed line) to supercritical one ($r_i\gtrsim
1/H_b$). Hereafter, it will evolve like the supercritical bubble
and contribute to SMPBHs. }
    \label{fig:Penrose3}
\end{figure}

As showed in Appendix-\ref{section_critical}, we also likely have
the bubbles with $M< M_{\text{cri}}\thickapprox{M_P^2\over H_b}$, but
$r\gtrsim {1\over H_b}$, see the right panel of \autoref{Vr} in
Appendix-\ref{section_critical}, such a bubble appeared at the
right side of barrier. Intuitively, it will also expand unbounded
and has a baby inflating universe inside its interior, like the
supercritical bubble.

In supercritical bubble case, initially $z<1$ (see
Appendix-\ref{section_critical}), we have \cite{Blau:1986cw} \be
\beta_{\text{Sch}}=\frac{1}{4}
\gamma^{4/3}H_b^{4/3}H_\sigma^{-1}\lf({2M\over
M_P^2}\rt)^{1/3}\lf({1-z^3\over z^2}\rt)>0. \ee Thus the Schwarzschild polar
angle is expanding, the initial evolution of bubble wall in the
Penrose diagram must move toward the right, see
\autoref{fig:Penrose1}. The corresponding mass $M_{\text{Sch}}$ is set by
$V(z=1)=0$, \be M_{\text{Sch}}= {M_P^2\over
2\gamma H_b}<M_{\text{cri}},\ee noting $\gamma^2=1+4H_\sigma^2/H_b^2>1$. However,
after $z>1$, we have $\beta_{\text{Sch}}<0$, which implies that the
Schwarzschild polar angle shrinks. In this case, the expansion of
bubble wall will move toward the left, and thus be eventually
hidden behind the horizon of a BH.


In view of this, the bubble with $M<
M_{\text{cri}}\thickapprox{M_P^2\over H_b}$ but $r\gtrsim {1\over H_b}$
is as if the subcritical bubble ($r< {1\over H_b}$) at the left
side of barrier \textsf{tunnel} to a larger bubble at its right
side, see \autoref{fig:Penrose3}. In this case, after inflation
ended the initial evolution of bubble is similar to that of
subcritical bubble for an observer outside bubble, but when $r$
arrived at $r_{\text{cls}}$ ($z_{\text{cls}}=({\gamma^2H_b^2M_P^2\over
2M})^{1/3}r_{\text{cls}}$, see Appendix-\ref{section_critical}), the
bubble will \textsf{up-tunnel} \footnote{Here, we use the term ``up-tunnel''
to distinguish it from the nucleation of bubbles.} to that
with $r_2\gtrsim 1/H_b$. Then it will expand unboundedly like the
supercritical bubble. The Penrose diagram is plotted in
\autoref{fig:Penrose3}.

In original Ref.\cite{Farhi:1989yr}, see also
\cite{Fischler:1990pk}, the probability of such a up-tunnelling
has been calculated. In the case of $M\gtrsim M_{\text{Sch}}$, we
have \footnote{ There is a uncertainty of sign in probability
exponent. Here, we follow original Ref.\cite{Farhi:1989yr}.} \be
P\sim e^{-{\pi M_P^2} \lf(r^2_2-r_{\text{cls}}^2\rt)}. \ee  This
probability is non-negligible when $r_2\longrightarrow r_{\text{cls}}$
($M \lesssim M_{\text{cri}}$). Thus even if the subcritical bubble is
unlikely to collapse to SMPBH, it might possibly contribute to SMPBH
by up-tunnelling to the supercritical bubble, which suggests a
novel channel to not only SMPBHs in our observable Universe but
also inflationary multiverse.

It can be expect that if $M\ll M_{\text{cri}}$, the up-tunnelling
probability will be exponentially low, however, the event number
might be not. It is significant to note \be \beta_{\text{dS}}=
\frac{1}{4}\gamma^{4/3}H_b^{4/3}H_\sigma^{-1}\lf({2M\over
M_P^2}\rt)^{1/3}\lf({1-\lf(2/\gamma^2-1\rt)z^3\over z^2}\rt),\ee and
$\beta_{\text{dS}}=0$ at $z=z_{\text{dS}}=\lf(2/\gamma^2-1\rt)^{-1/3}$,
$\beta_{\text{dS}}<0$ for
$\gamma^2<2$ (equivalently $H_b^2< 4H_{\sigma}^2$).
$\beta_{\text{dS}}<0$ ($>0$) suggests a shrinking (expanding) de Sitter
polar angle. The corresponding mass $M_{\text{dS}}$ is set by
$V(z_{\text{dS}})=0$, \be
M_{\text{dS}}=\lf(2-\gamma^2\rt) {M_P^2\over
2H_b}<M_{\text{cri}}.\ee

Thus in the case of $M\ll M_{\text{dS}}$, we have
\cite{Farhi:1989yr,Fischler:1990pk} \be P\sim e^{-{\pi
M_P^2}\lf({1\over H_b^2}-{4M^2\over M_P^4}\rt)}\sim e^{-{\pi
M_P^2\over H_b^2}}, \ee which is exponentially low, unless
${H_b^2\over M_P^2}\sim 1$. It is possible that we only observe
the ``universe" of ${\cal N}\sim 60$ efolds, while slow-roll
inflation might last ${\cal N}\gg 60$, so that the number of such
up-tunnelling bubbles is \be n\sim \lambda e^{3{\cal N}-{\pi
M_P^2\over H_b^2}}. \ee Thus we might have $n\gg 1$ for enough
large $\cal N$ \footnote{It seems that the efolds number of
inflation must be bounded by its dS entropy
\cite{Arkani-Hamed:2007ryv}, which implies ${\cal N}\lesssim {\pi
M_P^2\over H_i^2}<{\pi M_P^2\over H_b^2}$. However, in light of
recent resolution \cite{Almheiri:2020cfm} for the information
paradox of BH, such a bound for $\cal N$ is not necessary to exist
\cite{Piao:2023vgm}.}.



\section{Initial clustering of SMPBHs and its implication to nano-Hertz GWs}
\label{section_cluster}

\begin{figure}[htbp]
    \centering
    \includegraphics[width=6.9in]{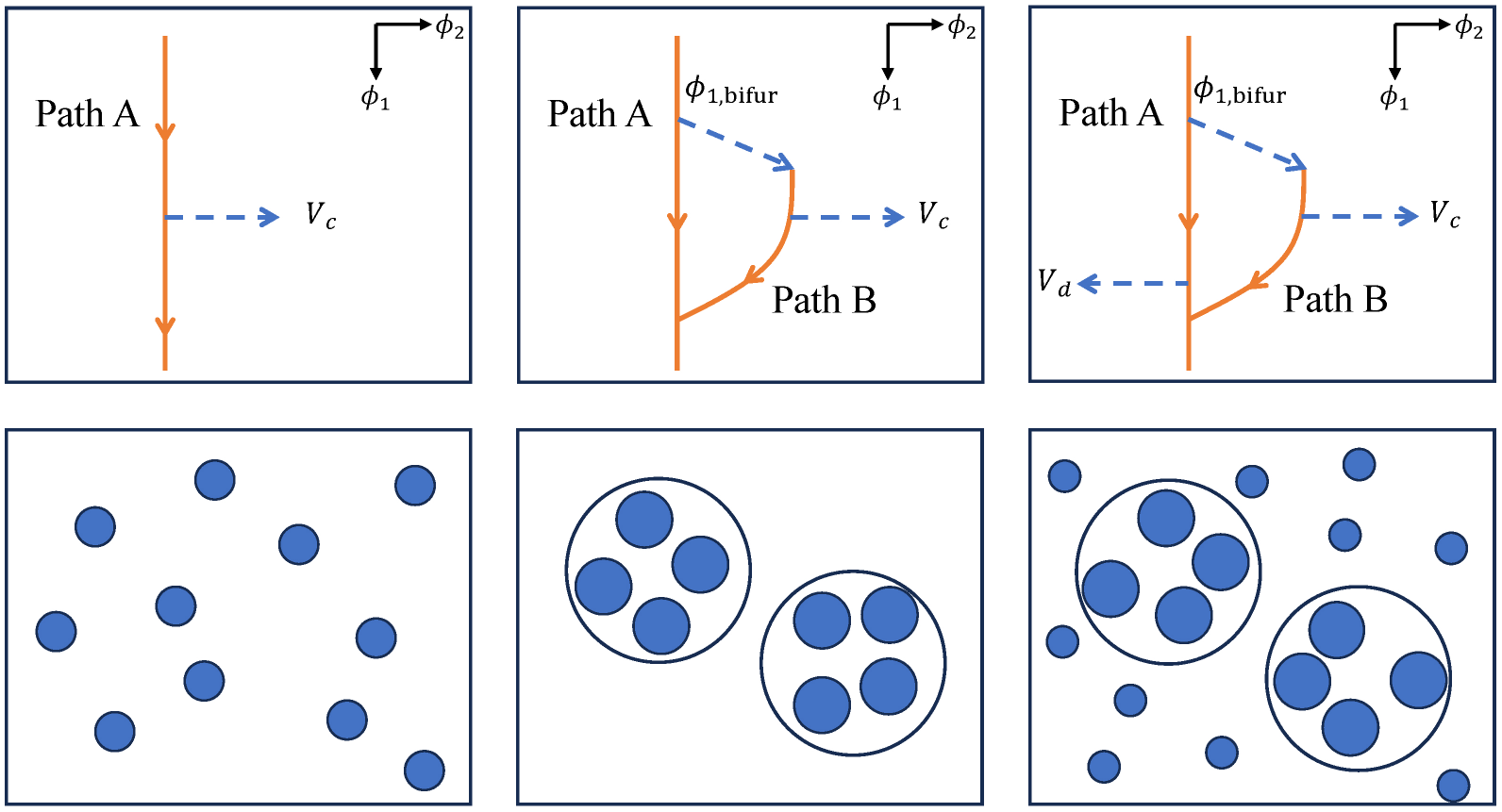}
\caption{\textbf{Initial clustering SMPBHs.} The left panel is the
sketch of model in \autoref{Two-FieldPotential}. In middle panel
the slow-roll path bifurcates at $\phi_{1,\text{bifur}}$ into Path A and
Path B while the bubbles with the vacuum $V_{c}$ spawned only in
the regions (bubbles) passing through Path B, which naturally
cause the clustering of SMPBHs. In right panel, Path A also
adjoins a different vacuum $V_d$. In middle and right panels, it
is required that before inflation ended both pathes converged.}
    \label{clusterPBH}
\end{figure}

It is well-known that the merger rate of PBHs binaries is
mass-suppressed, i.e. lower for PBHs with higher mass
(see e.g. Refs.\cite{Sasaki:2018dmp,Ali-Haimoud:2017rtz,Chen:2018czv,
Liu:2018ess,Kocsis:2017yty,Raidal:2018bbj,Huang:2023klk}).
Initial clustering PBHs can significantly magnify the merger rate
of PBH binaries \footnote{However, an excessively large PBH
abundance and intensive clustering can suppress the merger
rate, see e.g. \cite{Raidal:2017mfl,Kawasaki:2021zir}.} , which
thus might results in a detectable SGWB. In this section, we will
present a mechanism for the origin of initial clustering of
SMPBHs, and discuss the possibility that the merging of such
SMPBHs binaries explains recent NANOGrav signal.


In our model, see \autoref{clusterPBH}, similar to that in
Ref.\cite{Ding:2019tjk} based on the multi-stream inflation
\cite{Li:2009sp,Li:2009me} (see also
\cite{Wang:2010rs,Afshordi:2010wn}), initially the inflaton
$\phi_1$ slowly rolls along Path A, and the amplitude of
primordial perturbation is consistent with CMB
observations \footnote{Though based on $\Lambda$CDM model the
Planck collaboration has showed $n_s\approx 0.965$
\cite{Planck:2018jri}, $n_s=1$ ($n_s-1\sim {\cal O} (0.001)$) is
also observationally allowed
\cite{Ye:2020btb,Ye:2021nej,Jiang:2022uyg,Jiang:2022qlj} in light
of the pre-recombination resolution of recent Hubble tension, see
recent
Refs.\cite{Kallosh:2022ggf,Ye:2022efx,Jiang:2023bsz,DAmico:2021fhz,
Takahashi:2021bti,Fu:2023tfo} for its implication for inflation
models.}, but when $\phi_1$ rolls to $\phi_{1,\text{bifur}}$, the nucleation
of bubbles with $V_{b}\lesssim V_{\text{inf}}$ will happen \footnote{The
bubble will become supercritical after ${\cal N}\sim {\cal
O}(1)$.}. Thus at $\phi_{1,\text{bifur}}$ the path responsible for inflation
bifurcates into the Path A (the slow-roll inflation continued) and
Path B (the slow-roll inflation happened inside the interior of
bubble). It might be speculated that at certain point of Path B,
the bubbles with $V_{c}\lesssim V_{b}$ will spawn, and eventually
developed to SMPBHs. Thus if before inflation ended both pathes
converged, we will have a population of initial clustering SMPBHs.

It is interesting to estimate the effect of such initial
clustering PBHs on the merger rate of PBH binaries. Provided that
the probability at $\phi_{1,\text{bifur}}$ into Path B is
$\lambda_B$ (related to the nucleating rate of B-bubbles),
and in our observable Universe (all patches went through Path A or
B) $\rho_{\text{PBH}}$ is the energy density of PBHs while in Path
B that of PBHs is $\rho_{\text{PBH}}^B$, we have \be
f_{\text{PBH}}={\rho_{\text{PBH}}\over \rho_{\text{DM}}}=
\lambda_B{\rho_{\text{PBH}}^B\over \rho_{\text{DM}}}= \lambda_B
f_{\text{PBH}}^B,\label{fc}\ee where
$f_{\text{PBH}}^B={\rho_{\text{PBH}}^B\over \rho_{\text{DM}}}$,
and $\rho_{\text{DM}}$ is the energy density of dark matter. In
the case of $\lambda_B=1$, all patches responsible for the
observable Universe will come into Path B at
$\phi_{1,\text{bifur}}$. As a result, PBHs did not cluster.


The merger rate of PBHs which were effectively randomly distributed
in space and formed binaries in the early Universe is
\begin{equation}\label{Rcluster}
    R(t)=\frac{1}{2}\frac{n_T}{(1+z_
    \text{eq})^3}\frac{\text{d}P}{\text{d}t},
\end{equation}
where $n_T=f\rho_\text{eq}/\lf\langle m\rt\rangle$ is the total
average number density, $f={\rho_{\text{PBH}}/ \rho_m}\thickapprox
0.85f_{\text{PBH}}$ ($\rho_m$ is the matter density), $\rho_{\text{eq}}$ is the
energy density at matter-radiation equality $z=z_\text{eq}$.
In this section, we normalize the scale factor to unity at $z_\text{eq}$,
following the convention in \cite{Huang:2023klk}.

Generally, since the number density of SMPBHs are very low, the
effect of the Hubble expansion of background on the merger rate
can not be neglected. Taking the effect of cosmic expansion on the
comoving distance of PBH pairs into account, we have the
differential merger rate \cite{Huang:2023klk}
\begin{align} \label{eq:mergerate0}
    \mathcal{R}(m_i,m_j,t)={\text{d}R(t)\over \text{d}m_i\text{d}
    m_j}&\approx\frac{2.04\times10^8}{\text{Gpc}^3\text{yr}}
    f\lf(\frac{m_i}{M_\odot}\rt)^{-1}\lf(\frac{m_j}{M_\odot}\rt)^{-1}\lf(\frac{\lf\langle
    m\rt\rangle}{M_\odot}\rt)
    \notag \\
    &\times
    \lf(\frac{t}{t_0}\rt)^{-1}
    \psi(m_i)\psi(m_j)Y(m_i,m_j,t),
\end{align}
with \footnote{Note that the definition of $Y$ is different
from that in \cite{Huang:2023klk}, in which
$Y(m_i,m_j,t)\equiv\int_0^{X_\text{max}}\text{d}Xe^{-X}\mathcal{P}(\gamma_X)/X_\text{max}$.
Here, for later convenience, we adopt \autoref{eq:integrad0}.
}
\begin{align} \label{eq:integrad0}
Y(m_i,m_j,t)\equiv\int_0^{X_\text{max}}\text{d}Xe^{-X}\mathcal{P}(\gamma_X),
\end{align}
where
$\psi(m)\equiv\frac{m}{\rho_\text{\text{PBH}}}\frac{\text{d}n}{\text{d}m}$
is the probability distribution function of PBHs,
$\mathcal{P}(\gamma_X)\equiv\frac{\gamma^2_X}{(1+\gamma_X^2)^{3/2}}$
with
\begin{align}
    \gamma_X\approx10^{-3}f^{\frac{16}{21}}\lf(1+\frac{\sigma_{\text{eq}}^2}{f^2}\rt)^
    {-\frac{1}{2}}\lf(\frac{m_i}{M_\odot}\rt)^{\frac{1}{7}}\lf(\frac{m_j}
    {M_\odot}\rt)^{\frac{1}{7}}\lf(\frac{m_i+m_j}{M_\odot}\rt)^{\frac{12}{7}}
    \lf(\frac{\langle m\rangle}{M_\odot}\rt)^{-\frac{37}{21}}
    \lf(\frac{t}{t_0}\rt)^{\frac{1}{7}}X^{-\frac{37}{21}},
\end{align}
and $\sigma_\text{eq}^2$ the variance of density
perturbations of the rest of dark matter at $z_\text{eq}$
\cite{Ali-Haimoud:2017rtz}, and $X\equiv({x/ \Bar{x}})^3={4\pi}
n_Tx^3 /3$ (${\bar x}=({4\pi} n_T/3)^{-1/3}$, $x$ is the comoving
separation of the PBH pair), while
\begin{equation} \label{eq:Xmax}
X_\text{max}\equiv\lf(\frac{x_\text{max}}{\Bar{x}}\rt)^3=\frac{(m_i+m_j)n_T}
    {2\rho_\text{eq}}.
\end{equation}
The effect of cosmic expansion has been included in
$Y(m_i,m_j,t)$. It is expected that only when
\begin{equation}
    x<x_\text{max}=\lf(\frac{3}{8\pi}\cdot\frac{m_i+m_j}{\rho_\text{eq}}\rt)^{1/3},
\end{equation}
can the PBH pair come into being at redshift
$z=z_\text{dec}>z_\text{eq}$ with
$1+z_\text{dec}=(1+z_\text{eq})\lf(\frac{x_\text{max}}{x}\rt)^{3}$.

Here, with the initial clustering we can modify
\autoref{Rcluster} as \cite{Ding:2019tjk}
\begin{equation}\label{Rcluster2}
    R_{\text{cluster}}(t)=\frac{1}{2}\frac{n_T}{(1+z_\text{eq})^3}\lf.\frac{\text{d}P}
    {\text{d}t}\rt|_{f\to f/\lambda_B}.
\end{equation}
Thus \begin{align} \label{eq:mergerate}
    \mathcal{R}_\text{cluster}(m_i,m_j,t)&\approx\frac{2.04\times10^8}{\text{Gpc}^3\text{yr}}
    f\lf(\frac{m_i}{M_\odot}\rt)^{-1}\lf(\frac{m_j}{M_\odot}\rt)^{-1}\lf(\frac{\lf\langle
    m\rt\rangle}{M_\odot}\rt)
    \notag \\
    &\times
    \lf(\frac{t}{t_0}\rt)^{-1}
    \psi(m_i)\psi(m_j)Y_{\text{cluster}}(m_i,m_j,t),
\end{align}
with \footnote{In the case without initial clustering, the integral
\autoref{eq:integrad} has the analytical solution under the approximation of $e^{-X}\approx1$
\cite{Huang:2023klk}. However, in the case with initial
clustering, \autoref{eq:integrad} can be solved only numerically.}
\begin{align} \label{eq:integrad}
    Y_{\text{cluster}}(m_i,m_j,t)\equiv\lf.\int_0^{X_\text{max}}\text{d}Xe^{-X}\mathcal{P}(\gamma_X)\rt|
    _{f\to f/\lambda_B}.
\end{align}

It has been pointed out that SMPBHs with $M\approx 10^9M_\odot$
might explain SGWB detected recently by PTA experiments
\cite{Huang:2023chx,Depta:2023qst,Gouttenoire:2023nzr}
\footnote{It is also possible that this SGWB might be
inflationary GW
\cite{Vagnozzi:2023lwo,Vagnozzi:2020gtf,Benetti:2021uea,
Jiang:2023gfe,Piao:2004tq,Piao:2004jg,Liu:2011ns,Liu:2012ww,
Zhu:2023lbf,Cai:2016ldn,Cai:2015yza,Cai:2020qpu,Choudhury:2023kam,Oikonomou:2023bli,Datta:2023xpr},
see also other explanations
\cite{Wang:2023len,Ghosh:2023aum,Li:2023bxy,Liu:2023ymk,Niu:2023bsr,DiBari:2023upq,
Wang:2023sij,Du:2023qvj,Ye:2023xyr,Balaji:2023ehk,Zhang:2023nrs,
Li:2020cjj,An:2023jxf,Xiao:2023dbb,HosseiniMansoori:2023mqh,King:2023ayw,He:2023ado,Frosina:2023nxu,
Maji:2023fhv,Kawasaki:2023rfx,Kawai:2023nqs,Lozanov:2023rcd,Gangopadhyay:2023qjr,
Chen:2023zkb,Chen:2023uiz,Liu:2023hpw,Choudhury:2023fjs}.},
specially, NANOGrav signal. The spectrum of SGWB is the
incoherent superposition of GW emitted by merging SMPBH binaries
over the whole cosmic history \cite{LIGOScientific:2016fpe}
\begin{equation} \label{eq:GWspectrum}
    h^2\Omega_{\mathrm{GW}}(\nu)=\frac{\nu}{\rho_c/h^2}\int\frac{\mathrm{d}
    R(m_i,m_j,z)\mathrm{d}z}{(1+z)H(z)}\left. \frac{\mathrm{d}E_
    {\mathrm{GW}}(\nu')}{\mathrm{d}\nu'}\right|_{\nu'=(1+z)\nu},
\end{equation}
where $\rho_c/h^2$ is the Hubble-rescaled critical density at
present time, $\nu'$ is the redshifted sourced frequence. It is well
known that the GW power emitted in the coalescence of non-spinning
BH binary is \cite{Ajith:2007kx,Ajith:2009bn}
\begin{equation} \label{eq:dEdf}
    \frac{\mathrm{d}E_{\mathrm{GW}}(\nu)}
    {\mathrm{d}\nu}=\frac{\pi^{2/3}M_{\text{chrip}}^{5/3}}{3M_P^{4/3}} \left\{
    \begin{array}{ll}
         \nu^{-1/3}, \quad &\nu<\nu_1  \\
         \nu_1^{-1}\nu^{2/3}, \quad &\nu_1\le \nu<\nu_2 \\
         \nu_1^{-1}\nu_2^{-4/3}\left[\frac{\nu}{1+4\left( \frac{\nu-\nu_2}{\sigma}
         \right)^2}\right]^2,
         \quad &\nu_2\le \nu<\nu_3 \\
         0, \quad &\nu_3\le \nu
    \end{array}
    \right.
\end{equation}
where $M_{\text{chrip}}=(m_im_j)^{3/5}(m_i+m_j)^{-1/5}$ is the
chirp mass and the parameters $(\nu_1,\nu_2,\nu_3,\sigma)$ are set by
$M_{\text{tot}}$ and $\eta$, see Ref.\cite{Ajith:2007kx}.

\begin{figure}[h!]
\centering
\includegraphics[width=0.65\textwidth]{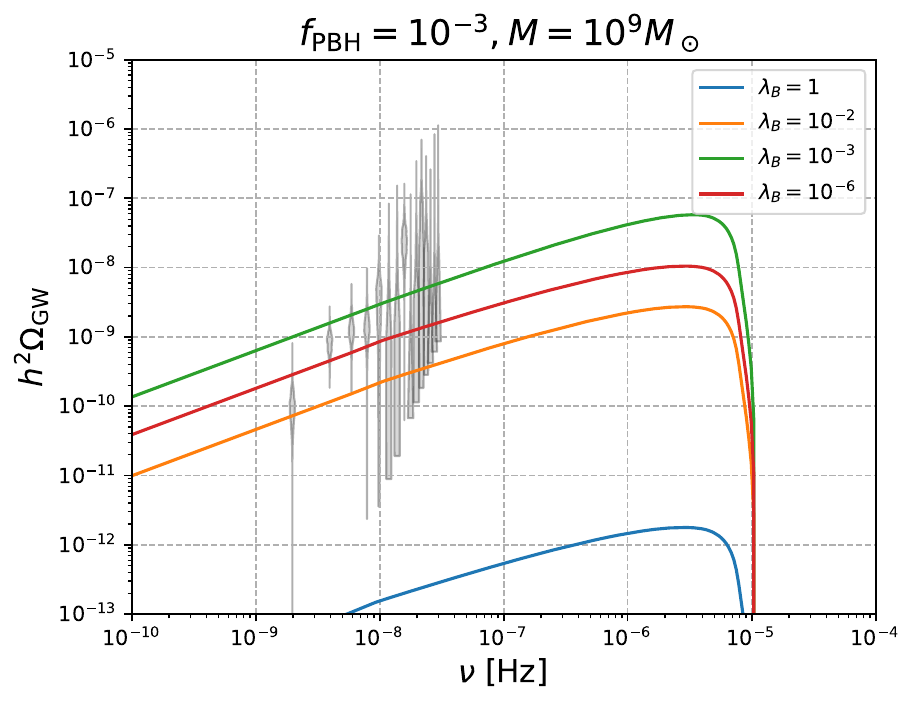}
\caption{\label{fig:PTA} The energy spectrum of the SGWB produced
by the monochromatic SMPBH binaries. The clustering level of PBHs
is depicted by the local density contrast $\lambda_B\ll 1$. It can
be seen that when $\lambda_B\sim 10^{-3}$ or $10^{-6}$, SMPBHs
with $10^9M_\odot$ can explain recent NANOGrav signal. In the case
of $\lambda_B\sim 10^{-4}$, the amplitude of SGWB is largest, when
$\lambda_B$ is lower than $10^{-4}$, local number density
of SMPBHs is larger, so that the pairing of SMPBHs binaries is
suppressed. }
\end{figure}

Assuming that the mass of SMPBHs is nearly monochromatic
\footnote{See Ref.\cite{Huang:2023klk} for the mass function of
SMPBHs sourced by the supercritical bubbles.}, we plot the
resulting SGWB in \autoref{fig:PTA}. According to current
cosmological and astrophysical constraints, we have
$f_\text{PBH}\lesssim0.001$ for $M\sim 10^9M_\odot$
\cite{Carr:2023tpt}. Thus it seems that SMPBHs must be initially
clustered to fit the NANOGrav data \cite{NANOGrav:2023gor}, our
model can just offer such a mechanism for the origin of initial
clustering SMPBHs.

However, the phenomenology of initial clustering of PBHs might be
richer than expected. It is also possible that before Path B
converged into Path A, at certain point of Path A the bubble with
$V_{d}\lesssim V_{\text{inf}}$ will also spawn, and sourced PBHs, see the
right panel of \autoref{clusterPBH}. Thus we have \be
f_{\text{PBH}}
=\lambda_Bf_{\text{PBH}}^B+(1-\lambda_B)f_\text{PBH}^A.\label{fd}\ee
In corresponding model, we will have not only some PBHs clusters
with $f_\text{PBH}^B$ but also almost uniformly distributed PBHs with
$f_\text{PBH}^A$. Thus the population of our multiverse PBHs can have
not only a multi-peaks mass spectrum, but also different
clustering levels.

In view of this, it is worth further exploring how the SMPBH
clustered and how well such SMPBHs can explain NANOGrav signal
(or, the possibility of detecting SMPBHs with PTA experiments). We
will perform the MCMC analysis with recent NANOGrav data
elsewhere.

\section{Conclusion} \label{section_discussion}


The supercritical bubbles ($r\gtrsim 1/H_b$) that nucleated during
slow-roll inflation can develop into SMPBHs, specially the mass of
SMPBHs can naturally have a peak at $10^5-10^{18}M_\odot$
\cite{Huang:2023chx}. In this paper, we further investigated
issues related with such SMPBHs.

It is possible that the subcritical bubbles ($r<1/H_b$) collapse
into the PBHs. Theoretically, such PBHs can be supermassive,
$M_{\text{PBH}}\gtrsim 10^5M_\odot$, only if slow-roll inflation happen
at a sufficiently low scale. However, after performing the
relativistic simulation for the collapsing of subcritical bubbles,
we found that large fluctuations of scalar field and spacetime
will be excited inevitably around the regions swept by the
collapsing bubble wall, which will substantially suppress the mass
of PBHs expected. In this sense, it seems unlikely for the
subcritical bubbles to collapse to SMPBHs.

However, before the subcritical bubble collapsed, it might have a
probability of up-tunnelling to the supercritical one ($r\gtrsim
1/H_b$), and thus contribute to SMPBH.
In seminal Refs.\cite{Farhi:1986ty,Farhi:1989yr}, the possibility
that the subcritical bubble up-tunnels to a baby inflating
universe has been explored. In multiverse PBHs scenario, slow-roll
inflation naturally offers a bank of such subcritical bubbles.
Thus inflating baby universe might also exist inside the interior
of bubble that is initially subcritical, which suggests that
inflationary multiverse might be more popular in such a story for
SMPBHs.


In our multiverse PBHs scenario, the rolling of inflaton might not
only be multiple-pathes, but also pass by multiple neighboring
vacua, so that the resulting multiverse PBHs would not only be
massive and supermassive with a multi-peaks mass spectrum, but
also cluster initially at different levels. This will bring richer
phenomenology than expected. As showed, the merging of initial
clustering SMPBHs is likely to explain recent NANOGrav signal. It
will be also interesting to study the implications of such
stellar-mass PBHs on LIGO/Virgo GWs e.g.\cite{He:2023yvl}.
Moreover, if different populations of PBHs with extreme mass-ratio
form, stellar-mass PBHs might be gravitationally bound to SMPBHs
($10^5 -10^9M_\odot$) in the early Universe, which together with
SMPBHs binaries contribute GWs sources for space-based detectors,
such LISA and Taiji (can hear the merger events of PBHs at higher
redshift $z\gtrsim 20$). Theoretically, the multi-peaks mass
spectrum and initial clustering of PBHs encoded the information of
not only slow-roll inflation but also the string vacua,
e.g.\cite{Douglas:2006es,Douglas:2019kus}, which thus might be an
prospective probe to new physics.

It finally might be worth mentioning that if the bubble that
nucleated during slow-roll inflation is responsible for SMPBH,
such bubble is more likely supercritical, thus the corresponding
observations hints for SMPBH will be also hints for the
\textsf{multiverse} scenario, i.e. well-known eternally inflating
multiverse \cite{Vilenkin:1983xq,Linde:1986fd}, see also
\cite{Guth:2007ng,Linde:2015edk}.


\section*{Acknowledgments}
YSP is supported by NSFC, No.12075246, National Key Research and
Development Program of China, No. 2021YFC2203004, and the
Fundamental Research Funds for the Central Universities. We thank
Yong Cai, Jun Zhang for discussions on SMPBHs, Heling Deng for
discussions on vacuum bubbles, Jun-Qian Jiang for
discussions on the merger rate of PBHs, and Eloy de Jong, Tiago
França, Bo-Xuan Ge, Pu-Xin Lin for helps on GRChombo code. We would also like
to thank the GRChombo Collaboration team
(http://www.grchombo.org/).

\appendix

\section{On supercritical bubbles}
\label{section_critical}

\subsection{The critical radius}

The metric inside the bubble is the de Sitter metric,
$g_{rr}=\lf(1-H_b^2r^2\rt)^{-1}$, while after inflation ended that
outside the bubble is approximately the Schwarzschild metric,
$g_{rr}=\lf(1-{2M\over M_P^2r}\rt)^{-1}$. In the thin-wall
approximation, the motion equation of bubble wall is
\cite{Blau:1986cw}\footnote{The mass \autoref{massparameter} of
bubble can be worked out with
$\beta_{\text{Sch}}^2=\lf(\beta_{\text{dS}}-2H_{\sigma}r\rt)^2$. }
\be \beta_{\text{dS}}-\beta_{\text{Sch}}=2H_{\sigma}r, \ee where
$H_\sigma=2\pi{\sigma\over M_P^2}$, with \be
|\beta_{\text{dS}}|=\sqrt{1+\lf({\text{d}r\over
\text{d}\tau}\rt)^2-H_b^2r^2},\quad\quad
|\beta_{\text{Sch}}|=\sqrt{1+\lf({\text{d}r\over
\text{d}\tau}\rt)^2-{2M\over M_P^2r}}.\ee It can be rewritten as
\be \lf({\text{d}r\over \text{d}\tau}\rt)^2 +V(r)=0,
\label{drdt}\ee where \be
V(r)=1-\gamma^{2/3}H_b^{2/3}\lf[{(1-z^3)^2\over
4z^4(1-1/\gamma^{2})}+{1\over z}\rt]\lf({2M\over
M_P^2}\rt)^{2/3},\label{V}\ee where
$\gamma^2=1+4H_\sigma^2/H_b^2$ and $z^3={\gamma^2H_b^2M_P^2\over
2M}r^3$.

\begin{figure}[tb]
   \subfigure[] {\includegraphics[width=3in]{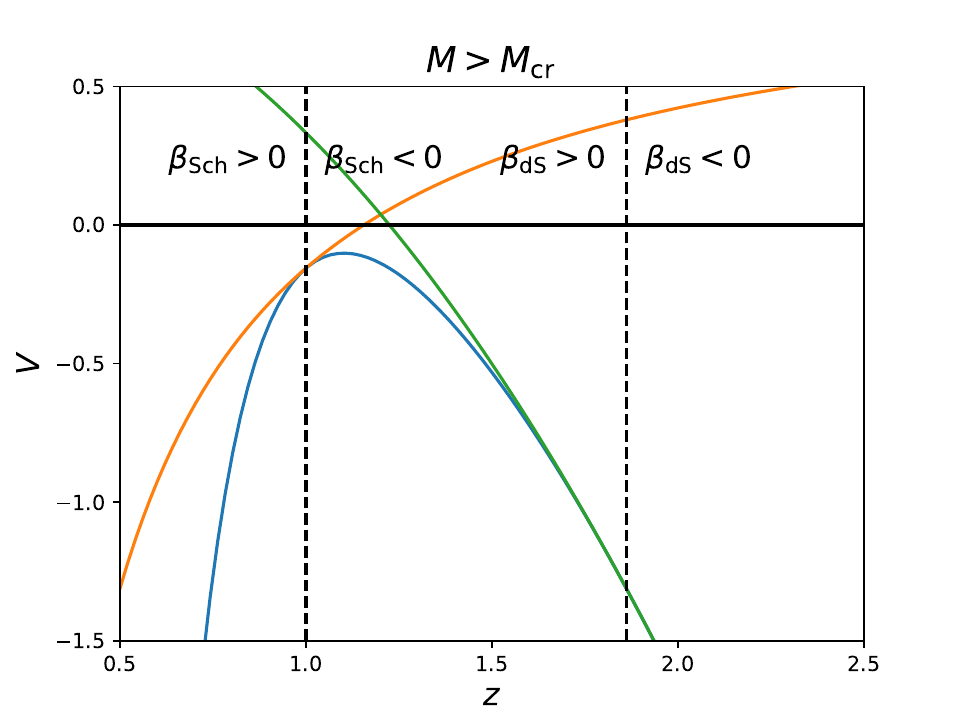}}
    \quad
   \subfigure[] {\includegraphics[width=3in]{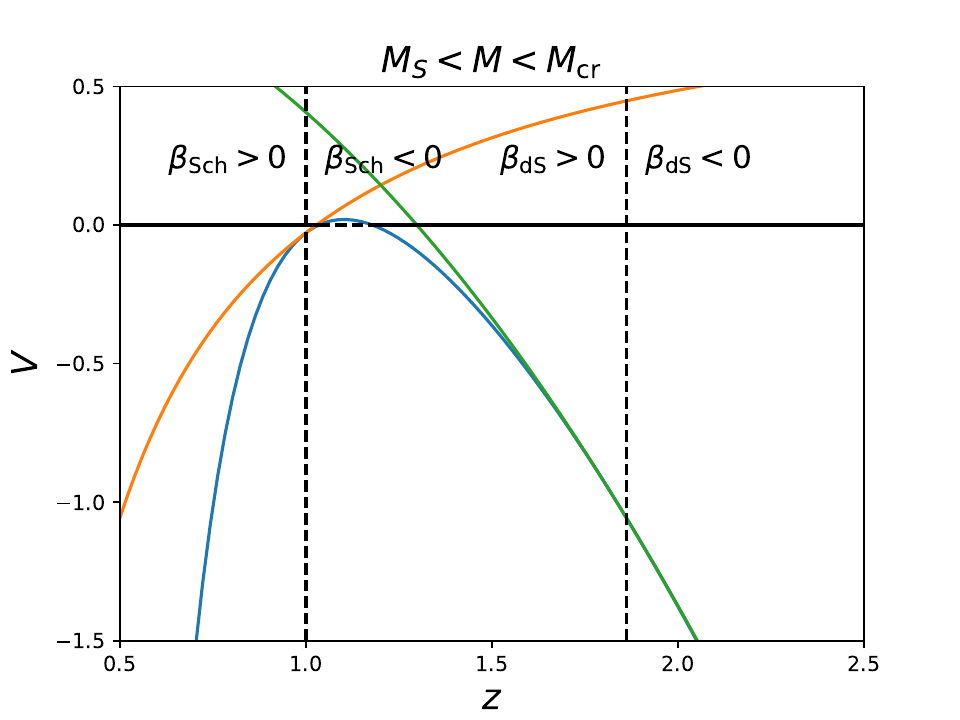}}
\caption{\textbf{Profile of the potential barriers \autoref{V} with
respect to $z$}. The motion of bubble wall is identical to that of
a particle in this potential (black solid line). $\beta_{\text{Sch}}$ and
$\beta_{D}$ intersect with the potential, the masses at
intersection points correspond to $M_{\text{Sch}}$ and $M_D$,
respectively. The Schwarzschild horizon ($r={2M\over M_P^2}$,
orange curve) outside bubble and the de Sitter horizon ($r={1\over
H_b}$, green curve) inside bubble are also showed. }
    \label{Vr}
\end{figure}

The evolution of bubble wall is controlled by $V(r)$, plotted in
\autoref{Vr}, which is a barrier and has a maximum $V_{\text{max}}$ at
$z=z_{\text{cri}}$ (equivalently $r=r_{\text{cri}}$). In light of \autoref{V},
$V_{\text{max}}-1\sim -M^{2/3}$ is smaller for a larger $M$. The
critical mass $M_{\text{cri}}$ (at which $V_{\text{max}}=0$) is \be M_{\text{cri}}=
{M_p^2\over 2\gamma H_b \lf[{(1-z_{\text{cri}}^3)^2\over
4z_{\text{cri}}^4(1-1/\gamma^{2})}+{1\over z_{\text{cri}}}\rt]^{3/2}}\thickapprox
{M_P^2\over H_b}, \ee for $H_b\gg H_\sigma$. Accordingly, we have
\be r_{\text{cri}}\sim {1\over H_b}. \ee

The mass of supercritical bubble is $M\gtrsim
M_{\text{cri}}\thickapprox{M_P^2\over H_b}$, thus the supercritical
bubble will unboundedly expand, see the left panel of
\autoref{Vr}, and contains a baby inflating universe inside its
interior, since $r\gtrsim r_{\text{cri}}\sim {1\over H_b}$. The
subcritical bubble has its mass $M< M_{\text{cri}}\thickapprox{M_P^2\over
H_b}$, thus it will inevitably collapse after $r$ meet the barrier
$V(r)$, see the right panel of \autoref{Vr}.

In Ref.\cite{Blau:1986cw,Farhi:1989yr}, be the bubble
supercritical or subcritical, both begin in ``singularity"
$(r=0)$. Here, both bubbles start to evolve only after inflation
ended, since they are \textsf{false-vacuum} bubble only after
that. It is significant that inflation seems naturally offer a
bank of such bubbles without initial singularity, see also
\autoref{fig:Penrose1}, \autoref{fig:Penrose2} and
\autoref{fig:Penrose3}.

In addition, it is also interesting to note that if $H_b\ll
H_\sigma$, we have $r^3_{\text{cri}}\thickapprox {M\over
H_\sigma^2M_P^2}$. Thus $M_{\text{cri}}\thickapprox M_P^2/H_\sigma$,
which suggests that the corresponding critical radius is \be
r_{\text{cri}}\sim {1\over H_\sigma}. \ee

\subsection{On the mass of PBHs sourced by supercritical bubbles}

In light of \autoref{drdt}, we have \be r\sim e^{H_b\tau}, \ee for
$r\gg 1/H_b$, since $V(r)\sim 1-H_b^2r^2\sim -H_b^2r^2$. Thus the
interior of supercritical bubble inflates, and the expansion is
far faster than its exterior ($a(t)\sim t^{1/2}$).

The mass of bubbles at $t_i$ (when inflation ended) is \be
M_{\text{bubble}}\simeq {4\pi\over 3}r^3_iV_b\sim{M_P^2\over
H_i}\lf({H_b\over H_i}\rt)^2e^{3{\cal N}_*}, \ee for negligible
$\sigma$, where the bubble nucleated at ${\cal N}_*$. Thus after
inflation just ended we have $M_{\text{bubble}}\propto e^{3{\cal N}_*}$. The
supercritical bubbles always expand, however, the internal
expansion of bubble will not engulf its exterior spacetime, and in
the angle of view of observers outside the bubble, the bubble wall
expanded comovingly with $a(t)$
\cite{Blau:1986cw,Garriga:1997ef} \footnote{This implies that a
wormhole will naturally start to come into being, which connected
the interior of bubble to the exterior, see e.g. Fig.4 in
Ref.\cite{Garriga:2015fdk}.}. The radius of bubble will cross the
cosmological horizon at $t_H$, where $r_{H}=a(t_H)r_i\sim t_{H}$,
we have \be t_H\sim {r^2_i\over t_i}=r_i^2 H_i\sim {1\over
H_i}e^{2{\cal N}_*}. \ee Thus \be M_{\text{bubble}}\simeq {4\pi\over
3}t_H^3V_b
\sim {M_P^2\over H_i}e^{2{\cal N}_*}\lf({H_b\over H_H}\rt)^2
\label{Mnotsuit}\ee noting ${H_H/ H_i}=e^{-2{\cal N}_*}$. Thus it
seems that the mass of PBHs sourced by such bubbles is
\autoref{Msuper} only for ${H_b\over H_H}\sim 1$. In the case of
${H_b\over H_H}\gg 1$, we have $M_{\text{bubble}}\gg {M_P^2\over
H_i}e^{2{\cal N}_*}$. However, the Schwarzschild radius of such
PBHs is $r_{\text{Sch}}={2M_{\text{bubble}}\over M_P^2}\sim r_{H}\lf({H_b\over
H_H}\rt)^2$, thus \be  r_{\text{Sch}}\gg r_H, \ee which is conflicted
with the causality.


In light of Refs.\cite{Farhi:1986ty,Blau:1986cw}, it will be
expected that when the supercritical bubble crossed into the
horizon of our observable Universe, it must be hidden behind the
horizon of a BH. In \autoref{fig:Penrose1}, this corresponds that
shortly after $t_H$ the bubble wall must cross through the
\textsf{``white hole"-like horizon} on the left side, which
suggests $r_H\sim {2M_{\text{PBH}}\over M_P^2}$, Thus we have \be
M_{\text{PBH}}\sim
{M_P^2 r_H}\simeq {M_P^2\over H_i}e^{2{\cal N}_*},\label{MptH}\ee
which is just \autoref{Msuper}. Thus the mass of corresponding
PBHs might be not that of the bubble but \textsf{the mass of
cosmological horizon} outside the bubble at $t_H$.


In a different angle of view, it looks like that the nucleation of
bubble during inflation excited large metric perturbations,
${\delta\rho/\rho}\gtrsim 1$, around the supercritical bubble,
which were stretched to the superhorizon scale with the comoving
expansion of bubble wall. Actually, \autoref{MptH} is just that of
PBHs caused by large inflationary perturbations,
e.g.\cite{Sasaki:2018dmp,Carr:2020gox}, \be M_{\text{PBH}}\simeq
\lf({k_*\over 4\times 10^6\text{Mpc}^{-1}}\rt)^{-2}M_\odot.\ee


\section{On our NR simulation}
\label{section_NR}


In our simulation for the collapse of subcritical bubbles, we used
the GRChombo package \cite{Clough:2015sqa,Andrade:2021rbd}.
In the context of 3+1 decomposition of NR, we have \be g_{00} =
-{\alpha}^2+{\beta}_i {\beta}^i ,\qquad g_{0i} = {\beta}_i ,\qquad
g_{ij} = {\gamma}_{ij}, \ee where $\alpha$ is the lapse,
${\beta}^i$ the shift vector and $\gamma_{ij}$ is the 3-metric on
the spacelike hypersurface with timelike unit normal
$n^\mu=(1/\alpha,-\beta^i/\alpha)$. The extrinsic curvature is
\begin{gather}
    {\cal K}_{ij}=-\frac{1}{2}\mathcal{L}_{\vec{n}}\gamma_{ij}.
\end{gather}
Thus we have
\begin{align}
    \tilde{\gamma}_{ij}=\chi\gamma_{ij},\quad \chi=\mathrm{det}(\gamma_{ij})^{-\frac{1}{3}}, \\
    \tilde{A}_{ij}=\chi({\cal K}_{ij}-\frac{1}{3}\gamma_{ij}{\cal K}),
    \quad {\cal K}=\gamma^{ij}{\cal K}_{ij}.
\end{align}

In order to make the evolution of spacetime and the matter inside
a well-posed Cauchy problem, the system of partial differential
equations should be explicitly hyperbolic-like. The corresponding
BSSN equations for NR are \cite{Baumgarte:1998te,Shibata:1995we}
\be
\partial_t\chi=\frac{2}{3}\chi\alpha {\cal K}-\frac{2}{3}\chi\partial_k\beta^k+\beta^k\partial_k\chi,
\ee \be
\partial_t\tilde{\gamma}_{ij}=-2\alpha\tilde{A}_{ij}+\tilde{\gamma}_{ik}\partial_j\beta^k+\tilde{\gamma}_{jk}\partial_i\beta^k-\frac{2}{3}\tilde{\gamma}_{ij}\partial_k\beta^k+\beta^k\partial_k\tilde{\gamma}_{ij},
\ee \be \label{eq:partialK} \partial_t{\cal K}=
-\gamma^{ij}D_iD_j\alpha+\alpha\lf(\tilde{A}_{ij}\tilde{A}^{ij}+\frac{1}{3}{\cal
K}^2\rt)+\beta^i\partial_i{\cal K}+{4\pi\over
M_P^2}\alpha(\rho+S), \ee \ba
\partial_t\tilde{A}_{ij}&=&\chi\left[-D_iD_j\alpha+\alpha\lf(R_{ij}-{8\pi\over M_P^2}\alpha S_{ij}\rt)\right]^{TF}+\alpha\lf({\cal K}\tilde{A}_{ij}-2\tilde{A}_{il}\tilde{A}^l_j\rt)\nonumber\\
&+&\tilde{A}_{ik}\partial_j\beta^k+\tilde{A}_{jk}\partial_i\beta^k-\frac{2}{3}\tilde{A}_{ij}\partial_k\beta^k+\beta^k\partial_k\tilde{A}_{ij},
\ea \ba
\partial_t\tilde{\Gamma}^i&=&2\alpha\left(\tilde{\Gamma}^i_{jk}\tilde{A}^{jk}-\frac{2}{3}\tilde{\gamma}^{ij}\partial_j{\cal K}-\frac{2}{3}\tilde{A}^{ij}\frac{\partial_j\chi}{\chi}\right)-2\tilde{A}^{ij}\partial_j\alpha+\beta^k\partial_k\tilde{\Gamma}^i+\tilde{\gamma}^{jk}\partial_j\partial_k\beta^i\nonumber\\
&+&\frac{1}{3}\tilde{\gamma}^{ij}\partial_j\partial_k\beta^k+\frac{2}{3}\tilde{\Gamma}^i\partial_k\beta^k-\tilde{\Gamma}^k\partial_k\beta^i-{16\pi\over
M_P^2}\alpha\tilde{\gamma}^{ij}S_j, \ea where $D_i$ are the
covariant derivative with respect to the spatial metric, and
$\tilde{\Gamma}^i=\tilde{\gamma}^{jk}\tilde{\Gamma}^i_{jk}$ is the
conformal connections.
The components of the energy tensor are
\begin{gather}
    \rho=n_an_bT^{ab},\quad S_i=-\gamma_{ia}n_bT^{ab},\quad S_{ij}=\gamma_{ia}\gamma_{jb}T^{ab},\quad S=\gamma^{ij}S_{ij}.
\end{gather}
In our simulation, the scalar fields, minimally coupled to
gravity, contribute $T_{ab}$ by
$T_{ab}=\nabla_a\phi\nabla_b\phi-\frac{1}{2}g_{ab}(\nabla_c\phi\nabla^c\phi+2V)$.


Here, $\alpha$ and $\beta^i$ are specified as unity and (0, 0, 0)
respectively on the initial hypersurface. We adopt the
moving-puncture gauge ($\eta\sim {{\cal O}(1)\over
2M_{\mathrm{ADM}}}$ and $\mu\sim {\cal O}(1)$)
\begin{align}
    \partial_t\alpha={}&-\mu\alpha {\cal K}+\beta^i\partial_i\alpha,\\
    \partial_t\beta^i={}&\frac{3}{4}B^i,\\
    \partial_tB^i={}&\partial_t\tilde{\Gamma}^i-\eta B^i.
\end{align}


The Hamiltonian and momentum constraints are
\begin{gather} \label{HamC}
{\cal
H}=\tilde{D}^2\chi-\frac{5}{4\chi}\tilde{\gamma}^{ij}\tilde{D}_i\chi\tilde{D}_j\chi+\frac{\chi}{2}\tilde{R}+\frac{1}{3}{\cal
K}^2-\frac{\chi^3}{2}\tilde{A}^{ij}\tilde{A}_{ij}-{8\pi\over
M_P^2}\rho=0,
\end{gather}
\begin{gather} \label{MomC}
{\cal
M}^{i}=\tilde{D}_j\tilde{A}^{ij}-\frac{3}{2}\chi^{-2/3}\tilde{\gamma}^{ij}\tilde{D}_j{\cal
K}-{8\pi\over M_P^2}\chi^{-2/5}S^i=0.
\end{gather}
Here, $\tilde{D}$ is the covariant derivative with respect to
$\tilde{\gamma}_{ij}$. Initially, we set
$\tilde{\gamma}_{ij}=\delta_{ij}$ and $\tilde{A}^{ij}=0$ on the
hyperslice. Accordingly, the momentum constraint \autoref{MomC} is
automatically satisfied, while the Hamiltonian constraint is
solved with the conformal factor $\chi$, which is iterated until
it converges to $\partial_t\chi={\cal H}=0$.







\end{document}